\documentclass[fleqn,usenatbib]{mnras}

\usepackage[T1]{fontenc}
\usepackage{ae,aecompl}

\usepackage{amsmath,amssymb}
\usepackage{graphicx}
\usepackage{xcolor}

\title[Stellar B field Interaction with a planet]{Numerical simulations of the interaction between the stellar magnetic field and a planet}

\author[Fabio De Colle et al.]{Fabio De Colle$^{1}$\thanks{E-mail: fabio@nucleares.unam.mx}, Douglas N.C. Lin$^{2,3}$, Chen Chen$^{4}$, Gongjie Li$^{5}$\\
$^{1}$Instituto de Ciencias Nucleares, Universidad Nacional Aut\'onoma de M\'exico, A. P. 70-543 04510 D. F. Mexico\\
$^{2}$Department of Astronomy and Astrophysics, University of California, Santa Cruz, CA 95064, USA\\
$^{3}$Institute for Advanced Studies, Tsinghua University, Beijing, China\\
$^{4}$School of Earth and Atmospheric Sciences, Georgia Institute of Technology, Atlanta, GA 30332, USA\\
$^{5}$Center for Relativistic Astrophysics, School of Physics, Georgia Institute of Technology, Atlanta, GA 30332, US
}

\date{Accepted XXX. Received YYY; in original form ZZZ}

\pubyear{2024}

\begin{document}
\label{firstpage}
\pagerange{\pageref{firstpage}--\pageref{lastpage}}
\maketitle

\begin{abstract}
Kepler and TESS observations led to the discovery of many close-in super Earths, including some with ultra-short orbital periods ($\lesssim 1$ day). During and shortly after their multi-Myr formation epoch, their GKM host stars generally have kilogauss magnetic fields which can exert torques on the orbits of nearby super- Earths. In this work, we examine one aspect of this interaction: the magnetic torque resulting from Alfvén-wing drag on non-corotating, non-magnetized planets engulfed by the host stars' stellar wind.  We compute the magnitude of this torque for a range of stellar magnetic field strengths, and planetary orbital velocities. We also model the planets' orbital evolution, taking into account for stellar spin down and magnetic field decay, and derive the boundaries within which ultra-short-period super-Earths can survive.
\end{abstract}

  
\section{Introduction} 
Among the thousands of exoplanets recently discovered, a significant portion orbits in close proximity to their host stars, with some even completing orbital periods in less than one day (i.e., ultra-short period planets; USP hereafter). The {\it in situ} formation of these planets poses a challenge, since the orbital distances of the USP planets are within the dust sublimation zone \citep{flock2019}. Various mechanisms have been proposed to explain the migration of these planets inward, including disk migration and in-situ formation followed by tidal decay \citep{lee2017}, dynamical migration due to interactions with planetary companions \citep{nagasawa2005, idalin2010, pu2019, petrovich2019}, as well as obliquity tides \citep{millholland2020}. However, the origins and orbital evolution of such celestial bodies remain largely unconstrained.

The discovery of these close-in exoplanets motivated us to study star-planet magnetic interactions to better understand their formation and orbital evolution. In particular, young T Tauri stars have radii 2-3 times that of the present-day Sun and surface magnetic fields of several thousand Gauss \citep{johns-krull2007}. These properties allow strong magnetic interactions between USP planets and their host stars in their infancy. When the star's rotation rate differs from the planetary orbital mean motion, and the magnetic field is in perfect co-rotation with the star, the differential motion induces an electric field that can generate a large current \citep{goldreich1969, laine2008}. 
The hypothesis of co-rotation remains valid as long as the planet is located inside the Alfvén radius, while it breaks down for larger radii.

The associated Lorentz force drives orbital evolution toward a synchronous state, similar to the effects of tidal interactions \citep{wei2024}. While these magnetic interactions are typically weak compared to gravity, cumulative effects of magnetic interactions can lead to orbital migration and heating of rocky planets \citep{kislyakova2017, kisliakova2018, kislyakova2020, noack2021, ahuir21a, ahuir21b, garcia23},  as well as gas giants \citep{laine2008, laine2012, strugarek2017, chyba2021}.

In this paper, we focus on unmagnetized rocky to Jupiter-sized planets orbiting around a magnetized host star. We neglect the effect of magnetic diffusion in the stellar atmosphere and interior \citep{laine2012}, a process which will be studied in a subsequent paper.  Despite the apparent simplicity of this configuration, several outstanding issues remain, including the structure of the induced field due to the interaction of the stellar magnetosphere with the planet and the possibility of reconnection of the induced field in the wake of the orbiting planets.  The efficiency of magnetic coupling between the planets and their host stars determines not only the rate of planetary orbital evolution but also the location of Ohmic heating and the intensity of potentially detectable electromagnetic radiation. 

In order to study these processes in detail, we adopt MHD simulations (\S\ref{sec:methods}) to calculate the resulting field structure and Lorentz torque (\S\ref{sec:fieldstructure}) on the planetary orbit, its effects on orbital migration (\S\ref{sec:migration}), and we also estimate these effects using a simplified analytical approach (\S\ref{sec:alfvenwings}).  The effects of magnetic diffusion in the planet and its host star will be analyzed and presented in a subsequent paper.  We summarize our results and discuss the radio signatures due to cyclotron radiation in \S\ref{sec:summary}.

\section{Simulations}
\label{sec:methods}

\subsection{Numerical method}

To study the interaction of the stellar magnetic field with a planet, we have carried out a set of three-dimensional (3D) simulations. We solve the resistive magnetohydrodynamics (RMHD) equations using the \emph{Mezcal} code \citep[][]{decolle2005, decolle2006, decolle2012}. The RMHD equations are:
\begin{eqnarray}
  \frac{\partial \rho}{\partial t}
    +\nabla\cdot(\rho \vec v)  &=& 0 \;,  \\
  \frac{\partial \rho \vec v}{\partial t}
    +\nabla\cdot
            (\rho \vec v \vec v - \vec B \vec B)&=&0 \;,\\
  \frac{\partial \vec B}{\partial t}
    +\nabla \times \vec E &=&0 \;.
\end {eqnarray}
In these equations, $\rho$ represents mass density, $\vec v$ is the velocity vector, $\vec B$ is the magnetic field normalized with respect to $\sqrt{4\pi}$, and $\vec{E}$ is the electric field. In all the equations, we take the speed of light as $c=1$. These equations describe the conservation of mass, momentum, and magnetic flux, respectively. Since the thermal pressure is negligible in the region between the planet and the star compared to the magnetic pressure, we do not explicitly integrate an energy equation or include the thermal pressure in the RMHD equations.

The electric field is related to the current density $\vec J$ (given by $ \nabla \times \vec{B}/(4\pi)$) through Ohm's law:
\begin{equation}
    \vec E = -\vec{v}\times \vec{B} + \eta \vec{J}\;,
    \label{eq:induction}
\end{equation}
where $\eta = \sigma^{-1}$ represents the electrical resistivity (or magnetic diffusivity), and $\sigma$ is the conductivity respectively.

To solve this system of equations, we use a second-order upwind scheme, which integrates the MHD equations using the HLLD method. The divergence of the magnetic field is maintained equal to zero (at machine precision) by using the constrained transport (CT) method \citep[e.g.,][]{toth2000}.

\subsection{Computational domain}

\begin{figure}
\centering
\includegraphics[width=0.45\textwidth]{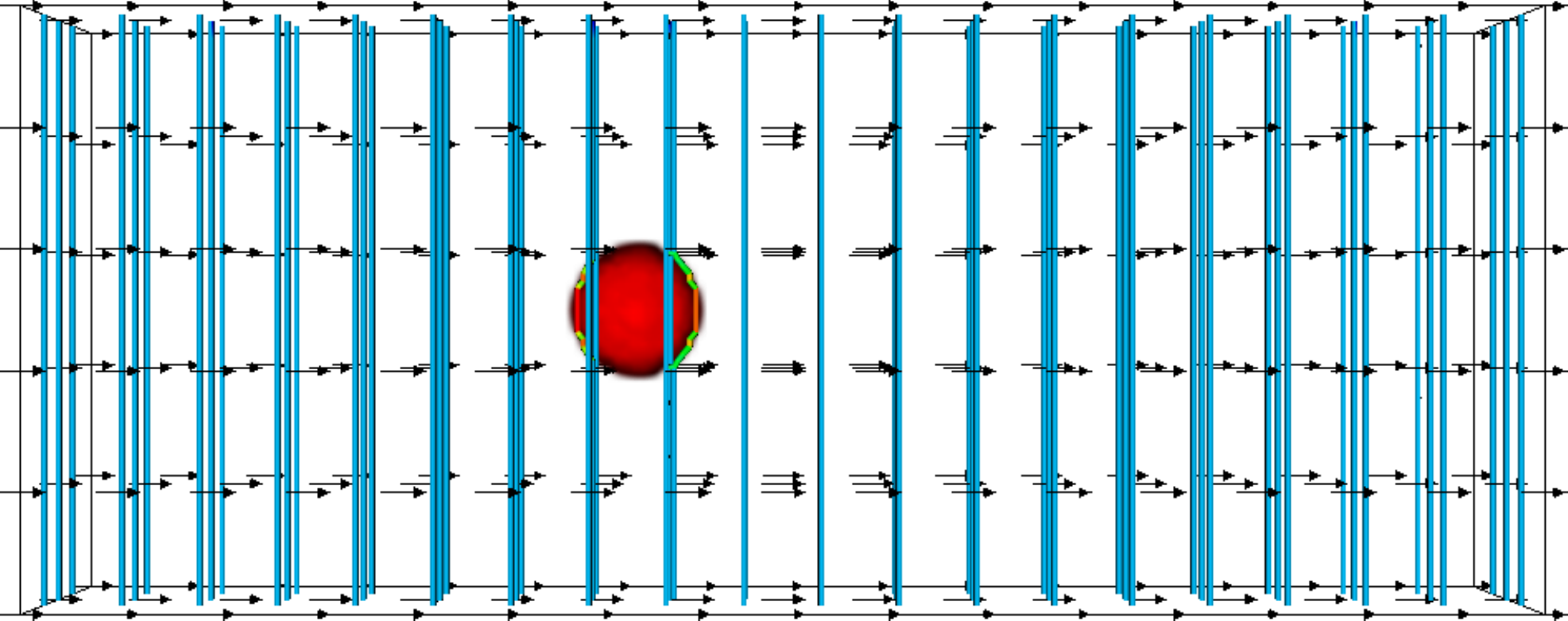}
\caption{Volume rendering showing the initial conditions of the simulations. The (x, y, z) coordinates are projected in the horizontal, vertical, and depth directions. The physical size of the computational domain (in normalized units) is: $(-1.5, 3.5), (-1, 1), (-0.5, 0.5)$ along the $x$, $y$ and $z$ axes respectively. The center of the planet (red circle) is located  at $(0.5,0,0)$, with a radius of $R_{\rm p}=0.2$. The magnetic field lines (blue lines) are initially vertical, with intensity $B_0$. The plasma density is set to $\rho=1$, with an initial velocity of $v_x=1$ (shown by the black arrows in the figure).}
\label{fig:init}
\end{figure}

The simulations employ a 3D Cartesian grid with a physical domain (in code units) of $(-1.5,3.5), (-1, 1), (-0.5, 0.5)$ along the $x$, $y$ and $z$ axes, respectively. The computational domain is discretized into $400\times160\times80$ cells, resulting in a resolution of $\Delta x=\Delta y=\Delta z= 0.0125$.

The initial conditions for the numerical simulations are shown in Figure \ref{fig:init}. These simulations of the interaction between the planet and the stellar magnetic field are done in the planet's reference frame, with all velocities normalized by the difference between the planet's orbital velocity and the velocity of the magnetic field. In this frame, the unmagnetized planet, with a radius of $R_{\rm p} =0.2$, is stationary and centered at $(0.5, 0, 0)$, while the magnetic field moves with a velocity of $v_{\rm p}=1$ in the positive $x$ direction (from left to the right) relative to the planet.
In all boundaries except the left boundary ($x=-1.5$), outflow boundary conditions are applied. For all variables except the magnetic field, these conditions are straightforward: variables are simply copied into the ghost cells. The magnetic field boundary condition is more complex: in the staggered mesh used in the CT method, tangential components (with respect to the boundary) are copied identically into the ghost cells, while normal components are inferred from the divergence-free condition $\nabla \cdot B = 0$:
\begin{eqnarray}
 \nabla \cdot B &=& \frac{b_{x,i+1/2,j,k}-b_{x,i-1/2,j,k}}{\Delta x} + \nonumber \\
&& \frac{b_{x,i,j+1/2,k}-b_{x,i,j-1/2,k}}{\Delta y} +  \nonumber \\ && \frac{b_{x,i,j,k+1/2}-b_{x,i,j,k-1/2}}{\Delta z} = 0
\end{eqnarray}
At the left boundary, inflow boundary conditions are applied by setting $b_{x,0,j,k} = b_{x,1,j,k}$, $b_{y,i=1,j\pm1/2,k}=B_0$, and $b_{z,i=1,j,k\pm1/2}=0$. This automatically guarantee that $\nabla \cdot B = 0$.
The velocity is fixed ($v_x=1$) at the top and bottom boundaries, for $|y| \ge 0.9$.

Additionally, we simplify the magnetic field structure by neglecting curvature effects and approximating the dipolar stellar magnetic field as an initial vertical magnetic field structure along the $y$ direction.

The initial stellar magnetic field is set as $\vec B=(0,B_0,0)$. All hydrodynamical codes present some degree of numerical viscosity, which can result in a loss of material from the planet. To suppress this numerical artifact, we enforce zero mass and momentum flux within the planet. This is done by: 1) setting a high density inside the planet ($\rho_{\rm p} = 10^{6}$), while the ambient medium has a density of $\rho_{\rm amb}=1$, and 2) imposing zero mass and momentum flux across cells with a density contrast greater than $10^5$. In this approach, the planet's density is treated as a passive scalar (used to track the planet's position), avoiding the need for additional equations and optimizing computing time. With this prescription, neither plasma nor magnetic fields are allowed to be advected through the planet. The resistivity in the ambient medium is set to $\eta_{\rm amb}=10^{-5}$.

As we show in the following section, the outcome of the numerical simulations depends primarly on three characteristic velocities: the Alfvén speed $c_{\rm B} =B_0/\sqrt{\rho_{\rm amb}}$ of the ambient medium, the diffusion velocity $v_{\rm d}=\eta_{\rm p}/R_{\rm p}$ of the magnetic field through the planet, and the relative velocity $v_{\rm p}$ between the unperturbed field and the planet.  Consequently, six different regimes arise from the possible combinations of these characteristic velocities.  In this paper,  we run numerical simulations for different values of $B_0$, specifically $B_0=10^{\alpha}$, with $\alpha=-3, -2.5, \dots, 3$. The diffusion of the magnetic field through the planet will be considered in a future study.

To verify numerical convergence, we carry out a set of numerical simulation varying the resolution and computational box size (see the Appendix).

\subsection{From code units to physical units}
\label{sec:codeunit}

Although the simulations use code units, the initial conditions and results can be rescaled to real astrophysical scenarios using a set of rescaling factors: $f_\rho$, $f_x$ and $f_v$, representing the rescaling factors for density, spatial dimensions, and velocity, respectively. These factors can be obtained through dimensional analysis, resulting in the following relationships:
\begin{eqnarray}
    f_\rho = \frac{\hat{\rho}}{\rho_{\rm amb}} = \hat{\rho}\;,\\
    f_x = \frac{\hat{R}_{\rm planet}}{R_{\rm p}} = 5 \hat{R}_{\rm planet}\;, \\
    f_v = \frac{\hat{v}_{\rm k}}{\hat{v}_{\rm p}} = v_{\rm k}\;,
\end{eqnarray}
where the $\hat{}$ symbol represents quantities in physical units. Thus, $\hat{\rho}$ is the density in the planetary environment, $\hat{R}_{\rm planet}$ is the planet radius, and $\hat{v}_{\rm k}$ is the difference between the planet's orbital velocity and the magnetic field's rotation velocity. The values used in the simulations are $\rho_{\rm amb}=1$, $R_{\rm p}=0.2$, and $v_{\rm p}=1$. The values of the magnetic and electric fields, as well as the time scale, can be determined using the rescaling factors: $f_B=(4 \pi f_\rho f_v^2)^{1/2}$, $f_E=f_B f_v$ and $f_t = f_x/f_v$. The characteristic velocity ratios discussed earlier are invariant under scaling, i.e. they remain the same in code units as well as in physical units. Thus, a single simulation, using different values of $\rho, R_{\rm planet}, v_k$, can describe the behavior of a wide range of physical systems governed by the same characteristic velocity ratios.

\section{Magnetic field structure} 
\label{sec:fieldstructure}

\begin{figure}
\centering
\includegraphics[width=0.45\textwidth]{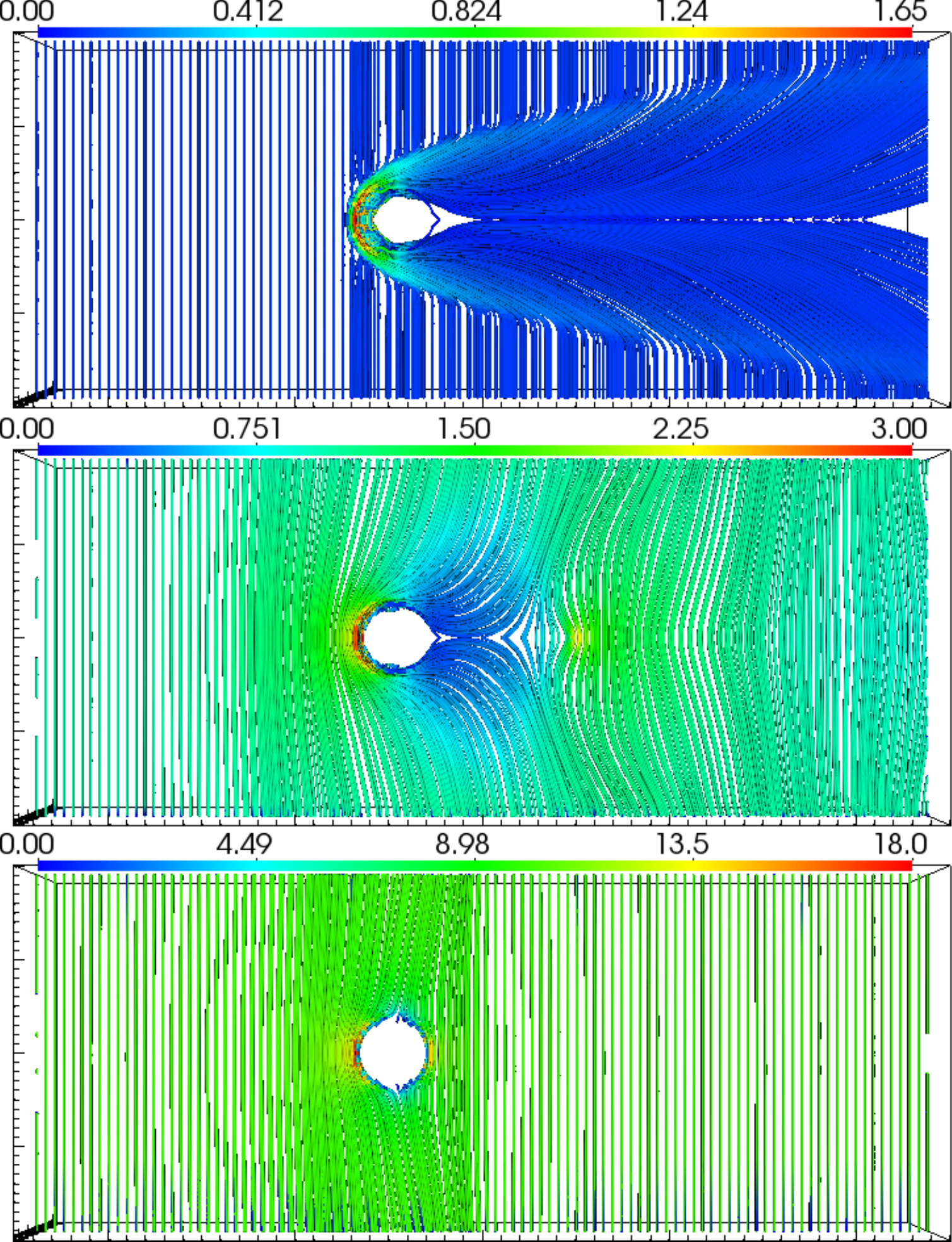}
\caption{Numerical simulations of the interaction between the planet and the stellar magnetosphere. The initial vertical magnetic field (see figure \ref{fig:init}) increases from top to bottom, with $B_0=0.1, 1, 10$ and corresponding Alfvén Mach numbers $M_{\rm A}=10, 1, 0.1$ in the top, middle, and bottom panels respectively. The planet's velocity is set to $v_{\rm p}=1$ in the three simulations. The resistivity is set as $\eta_{\rm p}=0$ inside the planet and $\eta_{\rm amb}=10^{-5}$ in the ambient medium. The figure shows a slice of the computational box, with streamlines generated in the plane $z=0$. The color bars represent the strength of the field lines.}
\label{fig:Bfield}
\end{figure}

In this paper, we focus on highly conductive planets with negligible magnetic diffusivity, a condition relevant to USPs with molten silicon surfaces. However, the surface of some slightly longer-period super-Earths may remain  solidified, leading to non-negligible magnetic diffusivity. While its contributions in equation (\ref{eq:induction}) are incorporated in the construction of our numerical method, we will analyze and present additional simulations exploring moderate to high magnetic diffusivity in a follow-up paper.

\subsection{The magnetic Mach number}
\label{sec:bmach}

Figure \ref{fig:Bfield} shows a three-dimensional representation of the magnetic field structure, displaying its evolution as the magnetic field strength increases (from top to bottom) {with values of} $B_0=0.1, 1, 10$. This figure highlights the strong dependence of the magnetic field structure on the ratio between the Alfvén speed, $c_{\rm B}$, and the relative velocity $v_{\rm p}$ of the planet with respect to the background magnetic field. The three panels correspond to specific values of $c_{\rm B}$ (0.1, 1, and 10 from top to bottom). Since $v_{\rm p}=1$ in all cases, these models correspond to magnetic Mach numbers $M_{\rm A}= v_{\rm p}/c_{\rm B}= (10, 1, 0.1)$. 

In the scenario in which the incoming flow is super-Alfvénic, i.e. $c_{\rm B} \ll v_{\rm p}$ (upper panel of Figure \ref{fig:Bfield}), the magnetic field lines are strongly stretched around the planet. In this case, magnetic reconnection occurring behind the planet limits the extent of stretching of the magnetic field lines. The location where most of the reconnection happens depends on both the magnetic field intensity and the resistivity of the environment, which is set to $\eta_{\rm amb}=10^{-5}$ in the ambient medium, in our simulations. 

As the planet moves through the stellar magnetic field, magnetic field lines accumulate in front of the planet, increasing both the magnetic pressure and tension in that region. The magnetic field intensity increases by a factor of $10-20$ in this region, reaching values $\sim 1-2$ (see upper panel of Figure \ref{fig:Bfield}). In steady state, the rate at which magnetic field lines accumulate  ahead of the planet balances the rate at which they are deflected sideways around the planet. As discussed in the following section, the planet's size, determining the rate of sideways expansion of the magnetic field, plays a crucial role in the system's long-term evolution. 

In the case $c_{\rm B} \sim v_{\rm p}$ (middle panel of Figure \ref{fig:Bfield}), magnetic field lines are modestly stretched around the planet. Magnetic reconnection occurs in this case just behind the planet, causing the magnetic field lines to return to a nearly vertical orientation beyond $x\gtrsim 2$. In front of the planet, the magnetic field intensity reaches values of $\sim 3$, i.e. about three times the original magnetic field  ($B_0=1$ in this case), while at the back of the planet the magnetic field intensity drops significantly to values $\ll 1$. 

In the limit $c_{\rm B} \gg v_{\rm p}$ (lower panel Figure \ref{fig:Bfield}), magnetic field lines remain nearly vertical throughout the computational box. In this scenario, the magnetic field reaches values $B \sim 18$ in front of the planet, i.e. only larger by a factor of $\sim 2$ with respect to the original unperturbed magnetic field in that region. Magnetic reconnection does not play a significant role in this particular case. Also, it is inferred that, although the magnetic field intensity in front of the planet is larger for increasing values of the unperturbed magnetic field, the ratio between the perturbed and unperturbed magnetic field becomes progressively smaller.

\begin{figure}
    \centering
    \includegraphics[width=0.45\textwidth]{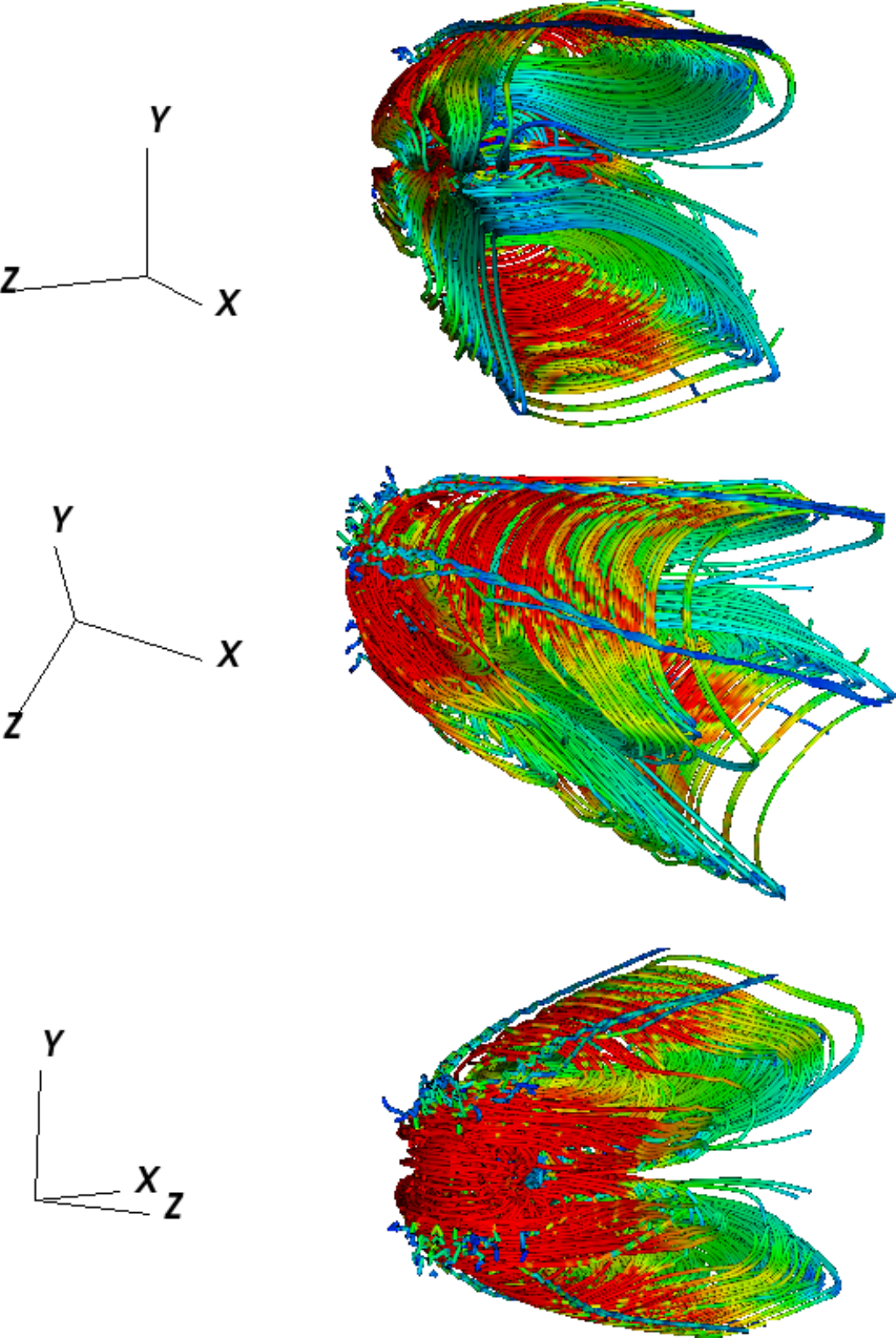}
    \caption{Current density ($\vec{J}=\nabla \times \vec{B}/(4\pi)$) streamlines for an initial magnetic field $B_0=0.1$ (the magnetic field streamlines corresponding to the same simulation are shown in the upper panel of Figure \ref{fig:Bfield}). The computational box is rotated as shown in the coordinate axes on the left. The color scale corresponds to the intensity of the current density, from blue (lower) to red (larger current density). As the magnetic field is nearly vertical, the current density is nearly zero on the left of the computational box (with respect to the $x$ axis). Downstream (with respect to the planet position), the current density streamlines form a complex structure, with cavities (with low current density) which start from the position of the planet and extend towards larger values of $y$ and $z$ (see upper right panel). The projection of the current density along the $x-y$ plane forms a structure which resembles the one obtained in analytical models (Figure \ref{fig:diff}). 
    }
    \label{fig:jfield}
\end{figure}

\subsection{Current density}
\label{sec:current}

The passage of the planet through the external field lines leads to a wake of induced fields (Figure \ref{fig:Bfield}) at an angle $\theta_{\rm A} \simeq {\rm tan}^{-1} M_{\rm A}$ \citep{neubauer1980}. In the Alfvén wing model, the associated current density 
\begin{equation}
  \vec{J}=\frac{\nabla \times \vec{B}}{4\pi}.
  \label{eq:current}
\end{equation}
is maintained across the wake. Circuit closure does not  necessarily occur on the stellar surface, as the current flow is regulated by the Alfvénic resistance, which allows for closure in regions such as the magnetosphere or ionosphere of the planet, rather than solely by magnetic diffusion on the stellar surface. We show the distribution of $J$ for a representative model (with $\eta_{\rm p}=0$ in the planet, $\eta_{\rm amb} =10^{-5}$, $B_0=0.1$, $v_{\rm p}=1$, and a corresponding $M_{\rm A} =10$) in Figure \ref{fig:jfield}.  
Intense current is induced in the leading hemisphere of the planet, where the perturbed field lines are markedly distorted as they wrap around the planet's surface\footnote{The small amplitude fluctuations upstream of the planet are a numerical artifact and are neglected in our analysis.}. The current also flows around the planet's surface but normal to the direction of the fields. 

The field strength and current in the trailing hemisphere are much weaker than those in the leading hemisphere.  Consequently, the perturbed field induces a Lorentz force against the planet's motion (\S\ref{sec:lorentztorque}). Downstream from the planet, both the perturbed field and current fan out from the planet's surface to its wake along a V-shaped Alfvén wing.

\begin{figure*}
    \centering
    \includegraphics[width=0.9\textwidth]{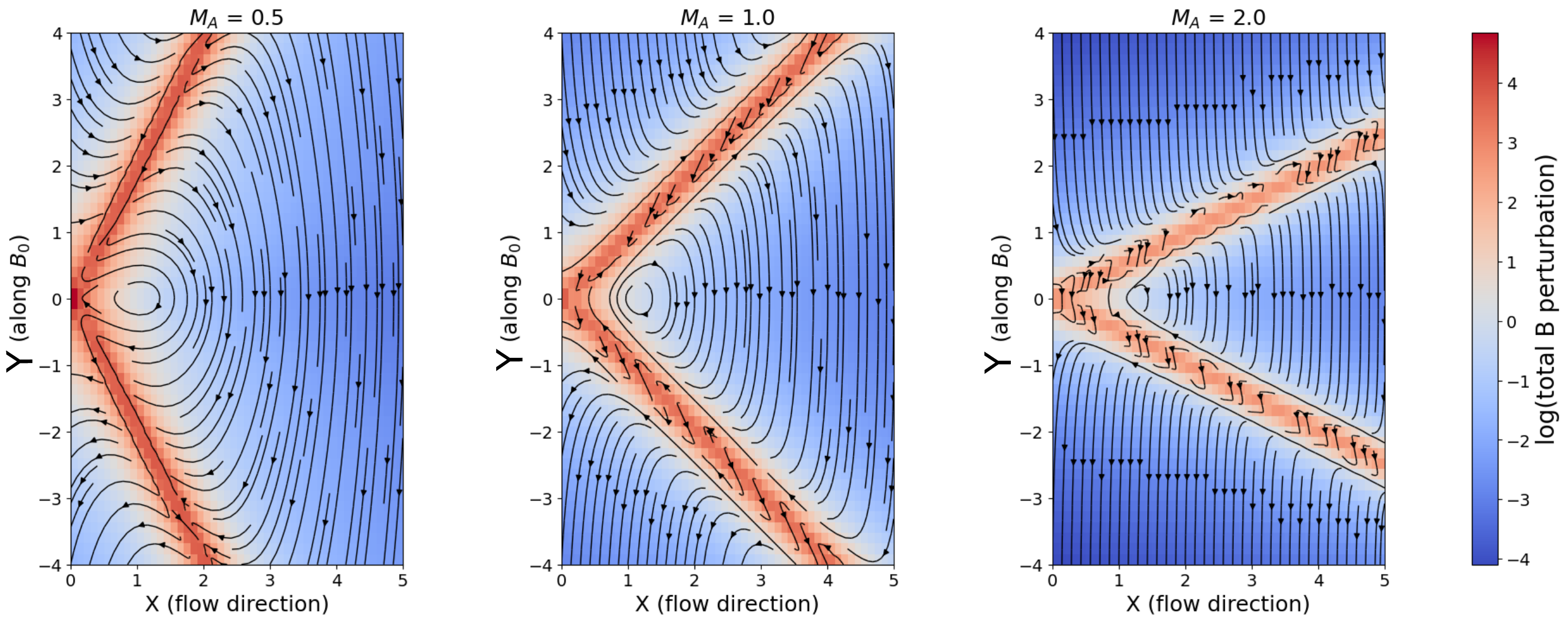}
    \caption{ 
    Alfv\'{e}nic perturbations for different Alfv\'{e}n Mach numbers ($M_A$) on the plane $Z = 0.1$, assuming the planet is a point source. Note that $Z\neq 0$, since the perturbation is zero exactly at $Z=0$ due to symmetry. From left to right, the $M_A$ values are 0.5, 1, and 2. The background magnetic field $B_0$ and current density $J_0$ are fixed at -1 and 10, respectively. The background color represents the total magnetic field perturbation in log scale. The streamlines show the direction of the total magnetic field (i.e., the perturbation plus the background field) on the $X-Y$ plane. 
    }
    \label{fig:diff}
\end{figure*}

\subsection{The Alfv\'{e}n-wing model}
\label{sec:alfvenwings}

In order to interpret these results, we adopt an idealized analytical MHD model by \citet{woodward1999}\footnote{Following \citet{woodward1999}, we keep SI units in this section.} to show the response of magnetic perturbation with respect to different magnitudes of Alfv\'{e}n Mach number ($M_{\rm A} \equiv v_{\rm p}/c_{\rm B} $) and current density ($\vec{J}$). This analytical model assumes the planet to be a conductor (with $\eta_{\rm p}=0$) as a point source at (0, 0, 0) and $\vec{J}$ has an initial amplitude $J_0 \delta (x) \delta(y) \delta(z)$ (where the symbol $\delta$ represents the $\delta$-function) in the $Z$ direction (in depth)\footnote{In principle, $J_0$ could be determined directly by considering the interaction of the magnetic field with a conductive sphere. This is done self-consistently in the simulations, where $J_0$ depends on the planet's velocity and the background magnetic field. In the Alfvén wing model, $J_0$ is treated as a free parameter, because this allows for an analytical treatment, instead of solving the full induction equation.
The analytical description then provides a qualitative description of the problem, which is validated by numerical simulations, that capture the full physical interaction.}. We assume the planet moves through an incompressible plasma, in a perpendicular ($X$) direction with respect to the background magnetic field (in the $Y$ direction) and the current (in the $Z$ direction). With the above definition of the current density, the magnetic field perturbation can be determined analytically. A detailed description of the model is given in equations (23-29) in \cite{woodward1999}. They are also shown below:
\begin{equation}
  B_x = - \frac{\mu_0 J_0}{M_A^2} (I_{B_x1} + I_{B_x2})\;,   \label{eq:bzy}
\end{equation}
\begin{equation}
  B_{y,z} = - \frac{\mu_0 J_0}{8\pi {M_{\rm A}} \kappa} \frac{\partial}{\partial y}
  \left(\frac{\partial}{\partial y}, \frac{\partial}{\partial z}\right) \log 
  \left[ \left(r - {\frac{M_{\rm A} x}{\kappa}} \right)^2 - {\frac{y^2}{\kappa^2}} \right] 
\end{equation}
where $\kappa \equiv \sqrt{1 + M_{\rm A}^2}$, $r\equiv \sqrt{x^2 + y^2 + z^2}$\;,
\begin{equation}
  I_{B_x1}  = \frac{M_{\rm A}^2}{4}\delta(z)[\delta(\zeta_+)+\delta(\zeta_-)]\;,
\end{equation}
\begin{align}
\begin{split}
I_{B_{x,2}} & = \frac{1}{4 \pi} \frac{\partial}{\partial y} \left(\frac{1}{r}\right)
- \frac{1}{8\pi M_{\rm A} \gamma^2} 
\{ 2\pi\delta(z)[\delta(\zeta_+)-\delta(\zeta_-)] \\
            & -\frac{1}{2\gamma}\frac{\partial^2}{\partial z^2}
            [{\rm log}(z^2 + {\zeta^2_+ \over \gamma^2})-{\rm log}
            (z^2+{\zeta^2_- \over \gamma^2})] \}\;, 
            \label{eq:i2}\\
\end{split}
\end{align}
and $\gamma \equiv \sqrt{1+M_{\rm A}^{-2}} = \kappa/M_{\rm A}$, $\zeta_{\pm}\equiv y \pm x/M_{\rm A}$.
In the above equation $B$ and $J$ normalized to the background $B_0$ and $J_0$, all the lengthscales are normalized with respect to $B_0/J_0$. The approximation of the planet as a point source is adequate for planets with $R_{\rm p} \ll B_0/J_0$.

Note that the wave angle $\zeta_\pm$ is related to $\theta_A \simeq {\rm tan}^{-1} M_{\rm A}$ whereas the induced field's strength is proportional to $J_0$ and declines as $M_{\rm A}^2$ in the limit $M_{\rm A} \gg 1$. This Alfvén wing model \citep{neubauer1980, woodward1999} is consistent with the simulation results shown in Figure \ref{fig:Bfield} and discussed in \S\ref{sec:current}. 
A similar comparison between the  Alfvén-wing model and the results of numerical simulations has been discussed by \citet{strugarek2015} (see their figure 3), for the case of dipolar and quadrupolar stellar magnetic fields, and for aligned and anti-aligned planetary magnetic fields.

Numerical solutions of equations (\ref{eq:bzy}-\ref{eq:i2}) show the magnetic perturbation for three different values of $M_{\rm A}$ (Figure \ref{fig:diff}). Similar to Figure (\ref{fig:Bfield}), the $x$-axis represents the flow direction, while the background magnetic field $B_0$ is pointing towards the negative $y$ axis. Magnetic field lines become more vertical as the Mach number increases (upper panels, from left to right panels) and are expected to be completely vertical for larger Mach numbers. The current flows along the $z$ direction (i.e. into the figure). In order to better illustrate the effect of $M_{\rm A}$, we fix $B_0=-1$ and $J_0=10$ respectively. As $M_{\rm A}$ increases, both the wake angle and the magnitude of perturbation decrease, which leads to more complex magnetic structure. For given values of $v_{\rm p}$ and $\rho$, large values of $M_{\rm A}$ correspond to a small value of $c_{\rm B}$ and a weak $B_0$.

\subsection{Effective conductance}
\label{sec:conductance}

The perturbed $E$ field is induced by the relative motion of the magnetized plasma and the planet, with a current density $\vec J$ (equation \ref{eq:current}). In the conventional unipolar-induction model \citep{goldreich1969, laine2012}, the current passes through a series circuit which includes both the planet and its host star.  In the Alfvén-wing model, there is no need for the current to pass through the footprint of the magnetic flux tube on the star \citep{neubauer1980}.

For a given E field, the Ohm's law equation (\ref{eq:induction}) indicates a dependence on the resistivity $\eta_{\rm p}$.  In principle, an arbitrarily large current can pass through a highly conductive planet.  However, in the limit of negligible resistance (i.e. $\eta_{\rm p} \rightarrow 0$) and sub-Alfvénic ($M_{\rm A} \leq 1$) flow, \citet{neubauer1980} showed that there is a maximum current $I_{\rm max} =\int J dA = 4E_0 R_{\rm p} \Sigma_A$ along the flux tube (containing the wake), where $E_0$ is the background electric field and $R_{\rm p}$ is the radius of the planet, due to the Alfv\'{e}n conductance across the wake
\begin{equation}
    \Sigma_{\rm A} = \frac{1}{\mu_0 c_{\rm B} (1 + {M_{\rm A}}^2 + 2 M_{\rm A} 
    {\rm sin} \theta)^{1/2}}.
\label{eq:al_cond}
\end{equation}
where $\theta$ represents the deviation of the direction of the planet relative to the direction perpendicular to the magnetic field (e.g., $\theta \sim 0$ for our case), and the expression is obtained under the assumption of a sub-Alfv\'{e}n flow model. The conductance has the dimension of time and should be replaced by the magnetic diffusion time $\sim R_{\rm p}^2/\eta_{\rm p}$ in the limit of large planetary resistance.  

\begin{figure}
\centering
\includegraphics[width=0.45\textwidth]{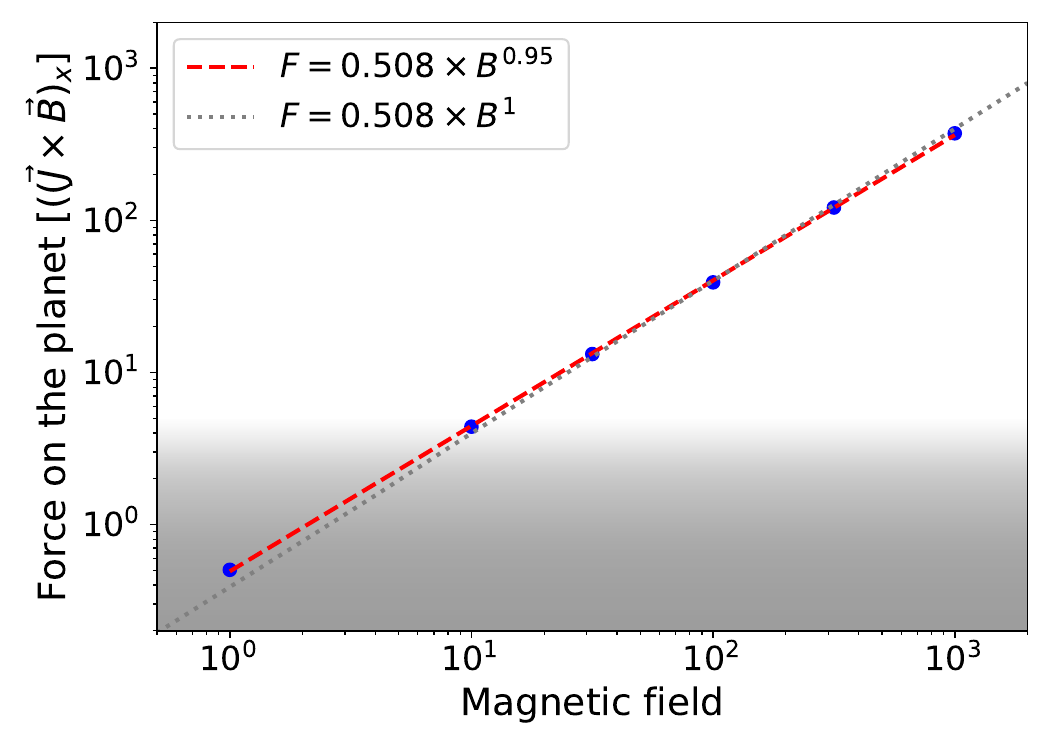}
\caption{Lorentz force acting on the planet due to its interaction with the stellar magnetic field, shown as a function of the stellar magnetic field strenght (blue dots). The best fit to the simulation results is given by $F= 0.508 \times B^{0.95}$ (red dashed line). The dotted grey line shows the analytical prediction from the Alfvén-wing model. The grey shaded area qualitatevely represents the region where ram pressure forces are expected to be important.
}
\label{fig:torque}
\end{figure}

\subsection{Calculation of the magnetic torque}
\label{sec:lorentztorque}

We compute the evolution of the torque exerted by the magnetic field on the planet by post-processing the results of the numerical simulations. The Lorentz force acting on the planet is defined as
\begin{equation}
 \vec{F} \equiv \int_V \vec{J}\times \vec{B} dV \;, 
\end{equation}
where the integral extends over the entire volume $V$ of the computational box, including the flux tube. In poorly conducting (small $\eta_{\rm p}$) planets, the field diffuses through the planet with modest distortion, with ${\vec J} \simeq \vec{v} \times \vec{B}/\eta_{\rm p}$, and the Lorentz force 
\begin{equation}
F \simeq J  B_0 R_{\rm p}^3 
\sim \frac{v_{\rm p} B_0^2 R_{\rm p}^3}{\eta_{\rm p}}\;.
\end{equation}

Around highly conductive planets, the magnetic diffusion time scale $R_{\rm p}^2 / \eta_{\rm p}$ becomes long compared with the characteristic dynamical timescale $a/ v_{\rm p}$.  According to the Alfvén-wing model, the current associated with the induced field can cross the wake with a maximum effective conductance in the flux tube of $\sim \Sigma_{\rm A}$ \citep{neubauer1980}. The maximum Lorentz force for highly conductive planets becomes 
\begin{equation}
    F_{\rm max} \simeq 2\alpha R_{\rm p} I_{\rm max} B_0 = { 8 \alpha M_{\rm A}
    B_0^2 R_{\rm p}^2 \over \mu_0   (1 + {M_{\rm A}}^2 + 2 M_{\rm A} 
    {\rm sin} \theta)^{1/2}}.  
    \label{eq:forcemax}
\end{equation}

We note that in the analytical calculations, we assume that Alfvén currents close on the surface of the planet. However, the effective area where the current closes is larger \citep{neubauer1980, strugarek2017}. Thus, we included a factor of $\alpha$ in the analytical expression (equation \ref{eq:forcemax}). This equation {\rm indicates} $F_{\rm max} \propto M_{\rm A} B_0^2 \propto B_0$ in the limit of large $B_0$ (when $M_{\rm A} \ll 1$) and $F_{\rm max} \propto B_0^2$ in the limit of small $B_0$ (and when $M_{\rm A} \gg 1$). Due to the symmetry of the problem considered, the net force along the $y$ and $z$ directions is zero. 

We compare this Alfvén-wing model for the highly conductive ($\eta_{\rm p} =0$) planets with the simulated results in Figure \ref{fig:torque}.
The magnitude of $F$ for various values of $B_0$ (bottom panel) is $\sim F_{\rm max}$. Moreover, the simulation shows that $F \propto B_0$ in the limit of large $B_0$ in accordance with the $F_{\rm max}$ dependence on $B_0$ shown in equation \ref{eq:forcemax}.
The net positive torque is produced mainly by the asymmetry between the magnetic field upstream ($-0.5<x<0$) and downstream ($0<x<0.5$) relative to the planet (see figure \ref{fig:res}). This headwind surplus leads to more intense dissipation on the leading hemisphere of the planet. Except for the small region around the planet, the magnetic field remains nearly vertical elsewhere. Figure \ref{fig:torque} also shows the region where ram pressure ($\rho_w v_w^2$), not considered in the rest of the paper, becomes larger that the Lorentz force.

We included the analytical results (computed by employing equation \ref{eq:forcemax}) as a dotted grey line in Figure \ref{fig:torque}.  
The effective interaction area is expected to be larger than the planet's surface (see, e.g., figure \ref{fig:jfield}). 
By comparing the analytical predictions with the numerical results, we scaled the analytical results by a factor of $\alpha = 1.6$ to account for the increase in the effective area. Once this rescaling factor is considered, the analytical predictions agree well with the numerical simulations. This scaling corresponds to an effective radius of 1.3 $R_p$, which is of the same order as the values found by \citet{strugarek2016}, who obtained values of $\sim 2-3$ for planetary magnetic fields aligned and anti-aligned with the stellar magnetic field. Our value is smaller due to the absence of a magnetosphere around the planet.


\section{Migration due to planet-star magnetic interaction}
\label{sec:migration}

In this section, we employ the results of the numerical simulations to determine the time evolution of the planet's orbit.

\subsection{The Alfvén-wing torque}
\label{sec:torque}

We consider a planet with a density $\rho_{\rm p}$, radius $R_{\rm p}$, (or mass $M_{\rm p}$), a circular orbit with a semi-major axis $a$, and an orbital angular frequency $\omega_{\rm p} = (G M_\star/a^3)^{1/2}$, orbiting its host at a distance $a$. The star has a mass $M_\star$, a magnetic field $B_\star$, and a radius $R_\star$, and it is spinning with a period $P_\star$ (or frequency $\Omega_\star=2 \pi/P_\star)$ and rotational axis parallel to the planet's orbital angular momentum vector.  For simplicity, we assume an aligned dipole stellar field such that its Lorentz force on the orbiting planet leads to a net torque only in the azimuthal direction\footnote{Other torque components can differ from zero if the orbital plane of the planet and the plane of symmetry of the stellar dipolar field, which typically coincides with the rotational plane of the star, do not coincide. In this case, the torque due to the magnetic field can be important in the evolution of the planet's orbital obliquity and eccentricity. This case will be considered in a future study.}. 
 
For a dipole field given by $B(a)=B_\star (R_\star/a)^3$, the net Alfvén-wing torque due to the Lorentz force on the planet, in physical units, is given by 
\begin{equation}
    \Gamma_{\rm A} = {\vec r} \times {\vec F} 
    \simeq 8 \alpha \Sigma_{\rm eff} (\Omega_\star - \omega_{\rm p}) 
    a^2 R_{\rm p} B_\star^2 (R_\star/a)^6\;,
    \label{eq:gammaa}
\end{equation}
where  the effective conductance is $\Sigma_{\rm eff}=\Sigma_{\rm A}$ (equation \ref{eq:al_cond}) in the limit $\eta_{\rm p}=0$.

In its tendency to reduce the relative motion between the field and the planet, the magnetic torque leads to planetary orbital evolution. In contrast to the null net torque in the radial and vertical directions, the non-vanishing Alfvén-wing torque along the azimuthal direction removes or feeds planet's orbital angular momentum 
\begin{equation}
     \frac{d(M_{\rm p} a v_{\rm k} (t))}{dt} 
     = \Gamma_{\rm A}\;,
    \label{eq:Revol}
\end{equation}
if the stellar magnetic field rotates slower/faster than the planet around the star\footnote{Here, we are considering the atmosphere of the star to be perfectly co-rotating, and we are neglecting the radial component of the stellar wind speed.} (i.e., if $a$ is smaller/larger than the corotation orbital radius $R_{\rm c} \equiv \Omega_\star^2 / G M_\star)^{1/3}$. The stellar magnetic field acts pushing the planet away from $R_{\rm c}$. 

In principle, the magnetic torque is applied in tandem with the tidal torque \citep{strugarek2017, ahuir2021, wei2024}.  These two contributions act in the same direction though their relative strength is a function of the planets orbital distance $a$ and their host stars' ages (see further discussions in \S\ref{sec:tidaltorque}). Before conducting comprehensive simulations of both effects, it is useful to assess the consequences of magnetic torque alone on the orbital evolution of hot Jupiters, close-in super Earths, and especially USPs, and to evaluate the minimum conditions required for their retention.

Since the planet's Keplerian speed $v_{\rm k}=(G M_\star/a)^{1/2}$, the orbital separation $a$ evolves as
\begin{equation}
     {d a \over dt} = {2 \Gamma_{\rm A} \over M_{\rm p}} 
     \left( {a \over G M_\star} \right)^{1/2}.
     \label{eq:evol}
\end{equation}
The right-hand term in this equation corresponds to the torque exerted by the magnetic field on the planet, which is determined directly from the results of the numerical simulations. The simulations are conducted in the planet's reference frame and $\Gamma_{\rm A}$ (or $\Sigma_{\rm eff}$) is primarily influenced by the ratio between Alfvén speed $c_{\rm B}$ and the planet's velocity ($v_{\rm p}= v_{\rm B}- v_{\rm k}= (\Omega_\star- \omega_{\rm p}) a$) relative to the rotation velocity of the magnetic field (\S\ref{sec:bmach}) which is assumed to be anchored on the star's surface. At each time, the new position of the planet is directly determined by integrating equation (\ref{eq:evol}). 

\subsection{Alfvén-wing induced planet migration}

For the torque, we interpolate logarithmically the values obtained by numerical simulations (see the bottom panel of Figure \ref{fig:torque}), which scales nearly as $\propto B$ (also see equation \ref{eq:forcemax}).  In the evaluation of the Alfvén speed $c_{\rm B}$ (used to select the appropriate range of simulations for interpolating the torque magnitude), the density of the ambient medium is calculated by considering the wind from the host star, i.e. 
\begin{equation} 
\rho_{\rm amb}=\dot{M}_{\rm w}/(4\pi a^2 v_{\rm w})
\label{eq:rhowind}
\end{equation}
with an outflowing speed $v_{\rm w}=500$ km s$^{-1}$ and a mass-loss rate $\dot{M}_{\rm w}$. During pre-main-sequence evolution, planets may be embedded in the residual protoplanetary and debris disks with a higher gas density (see further discussions in \S\ref{sec:ttauri}). Nevertheless, the strong field maintains relatively large values of $c_{\rm B}$ during the host star's infancy.

We integrate over a period of $10^9$ yrs, although most of the evolution in the planet's orbit takes place during the first 10$^8$ years. We neglect angular momentum changes in the star due to the magnetic torques on the planet, as the planet's angular momentum is typically several orders of magnitude smaller than that of young stars (see, e.g., \citealt{dobbsdixon2004}). Nevertheless it could slightly modify the angular momentum transferred to and from the planet. To qualitatively prescribe the decline over time of the star's magnetic field, wind mass-loss rates, and total angular momentum (due to magnetic braking), we assume the following:
\begin{eqnarray}
    B_\star(t) &=& B_{\rm i} \left(1+ \frac{t}{10^7 {\rm yrs}}\right)^{-1}\;,
    \label{eq:btime}\\
    \dot{M}_{\rm w}(t) &=& \dot{M}_{\rm i} \left(1+ \frac{t}{10^7 {\rm 
    yrs}}\right)^{-1.5}\;,
        \label{eq:mtime}
\\
        J_\star(t) &=& J_{\rm i} \left(1+ \frac{t}{10^7 {\rm yrs}}\right)^{-0.5}\;, 
    \label{eq:jtime}
\end{eqnarray}
so that they drop from their initial values $B_{\rm i}$, ${\dot M}_{\rm i}$, and $J_{\rm i}$ by factors of $10^2$, $10^3$, and 10 respectively, over 10$^9$ yrs \citep{skumanich1972, soderblom1993, stassun1999, dobbsdixon2004, folsom16, folsom18, wei2024}.  

We consider a range of initial values $B_{\rm i} = 1 - 10$ kG, and $ \dot{M}_{\rm i} =10^{-11}-5 \times 10^{-10}$ M$_\odot$ yr$^{-1}$ (see discussions in \S\ref{sec:initial}). We adopt a stellar mass $M_\star = 1 M_\odot$. In order to take into account the pre-main-sequence contraction, we also prescribe the stellar radius as
\begin{equation}
    R_\star (t) = {R_{\rm pms} \over (1 + t/10^7 {\rm yrs})^2} + 
    {R_{\rm ms} (t/10^7 {\rm yrs})^2 \over (1+ t/10^7 {\rm yrs})^2 }\;,
    \label{eq:rtime}
\end{equation}
where $R_{\rm pms} = 2R_\odot$ and $R_{\rm ms}=1 R_\odot$. The spin frequency and rotational period can be obtained using a constant coefficient of moment of inertia $\gamma_{\rm rot} \sim 0.2$ so that
\begin{equation}
    \Omega_\star= {J_\star \over \gamma_{\rm rot} M_\star R_\star^2}
    = \Omega_{\rm i} {R_{\rm pms}^2 \over R_\star^2} \left(1+ \frac{t}{10^7 {\rm yrs}}\right)^{-0.5}
\label{eq:omegatime}
\end{equation}
where $\Omega_{\rm i} = J_{\rm i} /\gamma_{\rm rot} M_\star R_{\rm pms}^2$. We adopt an initial spin frequency $\Omega_{\rm i} = 2 \pi/P_{\rm i}$ in the initial period range $P_{\rm i} = 1-14$ days (see below). With this prescription, we can account for the early increase of $\Omega_\star$ during the star's pre-main-sequence contraction and its later decline on the main sequence. Finally, we consider planets with a radius range $R_{\rm p}=0.1-10 R_\oplus$, a density range $1-5$ g cm$^{-3}$ and an initial orbital period range $P_{\rm p} =1-7$ days. 

The Alfvén wing model is only applicable inside the Alfvén radius, where the magnetic pressure is balanced by the ram pressure of the wind, i.e. for $a< R_{\rm A}$, where $R_{\rm A}$ is the Alfvén radius, defined implicitly by the relation $B (R_{\rm A})^2 = \rho (R_{\rm A}) v_{\rm w}^2$. 
Using this equality, along with equations (\ref{eq:btime}), (\ref{eq:mtime}), and (\ref{eq:rtime}), we find
\begin{equation}
 R_A = 
\left(\frac{4 \pi B_{\rm i}^2  R_\star(t)^6}{\dot{M}_{\rm i}  v_{\rm w}}\right)^{1/4} \left(1+ \frac{t}{10^7 {\rm yrs}}\right)^{-1/8} \nonumber
\end{equation}
\begin{equation}
= 174 R_\odot B_{\rm i,10}^{1/2} R_{\star,1}(t)^{3/2} (\dot{M}_{\rm i,11} v_{\rm w,7})^{-1/4} \left(1+ \frac{t}{10^7 {\rm yrs}}\right)^{-1/8}\;.
\end{equation}
In this equation, the initial magnetic field, wind mass-loss rate, and stellar wind are normalized with respect to 10 kG, 10$^{-11}$ M$_\odot$ yr$^{-1}$ and 100 km s$^{-1}$, respectively. The stellar wind velocity is assumed to be a constant as a function of time. Considering a varying stellar velocity would not substantially change the result, as $R_A$ depends weakly on $v_w$ (as $v_w^{-1/4}$), and the velocity is not expected to change substantially with time (as it remains of order of the escape velocity, see, e.g., \citealt{johnstone2015}). The stellar radius is normalized with respect to one solar radius. From equation (\ref{eq:rtime}), we get $1 R_\odot<R_\star(t)< 2 R_\odot$. Therefore, we conclude that in all the models we have computed, the Alfvén radius $R_A$ is much larger than the orbital separations considered.  

\subsection{Initial magnitudes of model parameters}
\label{sec:initial}

The above models are mostly relevant for hot Jupiters and close-in super Earth planets. These planets (especially those in mean-motion-resonance chains) may have emerged at large distances from the host stars, undergone disk migration to, and stalled near the proximity of their present-day orbits \citep{goldreich1980, lin1986, ward1997, ida2004, liu2017, Schoonenberg2019}.  Their initial staging post is determined by their magnetic and tidal interaction with their host stars \citep{lin1996}.

Many T Tauri stars have well-measured magnetic fields \citep{johns-krull2007} and show evidence of magnetospheric accretion \citep{bouvier2007}. The radius of the magnetospheric cavity, $R_{\rm cavity}$, in accretion disks (including protoplanetary disks) is estimated as 
\citep{ghosh1979, frank1992}
\begin{equation}
    {R_{\rm cavity} \over R_{\rm pms} }
    \sim \left( {B_\star ^4 R_{\rm pms}^5 \over G M_\star {\dot M}_{\rm disk}^2} \right)^{1/7}.
\end{equation}
For typical values of the disk accretion rate, ${\dot M}_{\rm disk} \simeq 10^{-8} M_\odot$ yr$^{-1}$ \citep{hartmann1998}, the cavity ratio is approximately 
\begin{equation}
    {R_{\rm cavity} \over R_{\rm pms} }
    \simeq 7 \left( {B_\star \over 10^3 {\rm G}} \right)^{4/7}     \left({10^{-8} M_\odot \over \dot{M}_{\rm disk} {\rm yr} } \right)^{2/7} 
\label{eq:cavitypms}
\end{equation}
around a solar-mass T Tauri star with $R_{\rm pms}=2 R_\odot$.  The differential angular frequency between the stellar spin $\Omega_\star$ and the Keplerian frequency $\omega_{\rm k} (R_{\rm cavity})$ induces a torque that drives $\Omega_\star$ toward synchronization with $\omega_{\rm k} (R_{\rm cavity})$.  In this state, $R_{\rm c} \simeq R_{\rm cavity}$ with $P_\star \sim$ a few days \citep{koenigl1991}. The observed $P_\star$ for T Tauri stars spans an order of magnitude \citep{stassun1999, broeg2006}, suggesting a wide dispersion in $B_\star$ and ${\dot M}_{\rm disk}$ \citep{serna2024}. 

The direction and speed of planetary migration due to planet-disk tidal interaction depend on the magnitude and gradient of the disk gas surface density distribution \citep{lin1986, paardekooper2010, paardekooper2011, chen2020}. Embedded in their natal protoplanetary disks, hot Jupiters' type II migration is halted upon entering the magnetospheric cavity \citep{lin1996}, while super-Earths' type I migration is interrupted by steep surface density gradients at the outer boundary \citep{liu2017}. However, due to magnetic and tidal interaction with their host stars, planets continue to migrate, albeit more slowly. In a synchronized state where $R_{\rm c} \simeq R_{\rm cavity}$, planets inside the cavity undergo orbital decay, while those outside experience orbital expansion. The gas density in both the magnetospheric cavity and the disk is much higher than $\rho_{\rm w}$, which generally suppresses $c_{\rm B}$ and increases $M_{\rm A}$. However,  in the limit $M_{\rm A} \gg 1$, $\Sigma_{\rm eff}$ and $\Gamma_{\rm A}$ remain independent of $M_{\rm A}$ (\S\ref{sec:lorentztorque}). With an appropriate determination of $\Sigma_{\rm eff}$, equation (\ref{eq:gammaa}) determines the survivability of protoplanets, particularly hot Jupiters, inside the magnetospheric cavity. 

\begin{figure}
\centering
\includegraphics[width=0.45\textwidth]{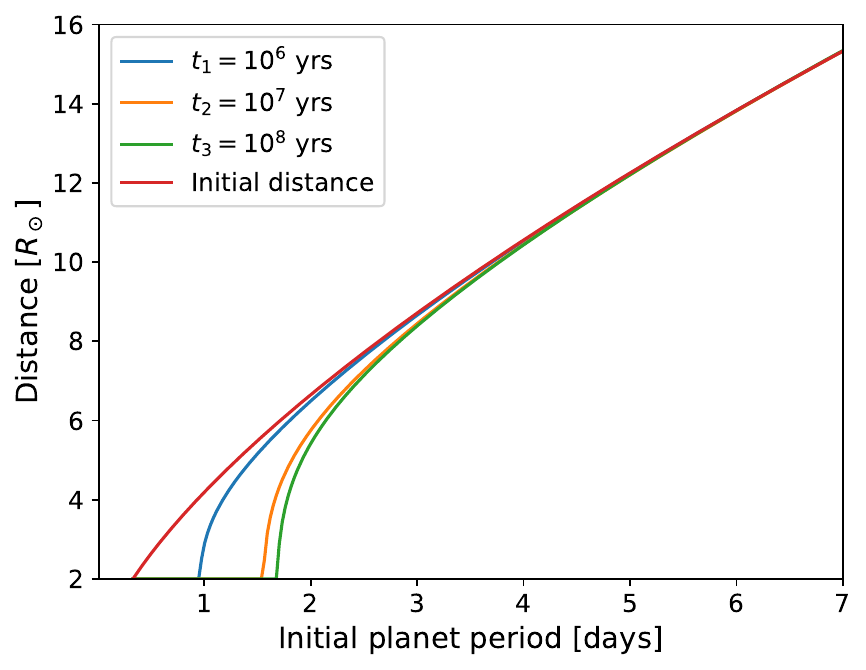}
\caption{Final orbital radius of a planet as a function of its initial orbital period, at different times (from $t=0$ yrs, red line, to $t=10^8$ yrs, green line). The planet's period evolves due to the Alfvén wing torque produced by the interaction between the stellar magnetic field and the planet. We consider a star with an initial magnetic field strength of 10 kG and an initial rotation period of 7 days. The star hosts a wind with an initial mass-loss rate of $5\times 10^{-10}$ M$_\odot$ yr$^{-1}$ and a velocity of 500 km s$^{-1}$. The planet has a radius twice that of Earth and a mass density of 4 g cm$^{-3}$. The evolution of stellar magnetic field, mass-loss rates, and spin angular momentum follow equations \ref{eq:btime}-\ref{eq:jtime}.}
\label{fig:orbvspla}
\end{figure}

\begin{figure*}
\centering
\includegraphics[width=\textwidth]{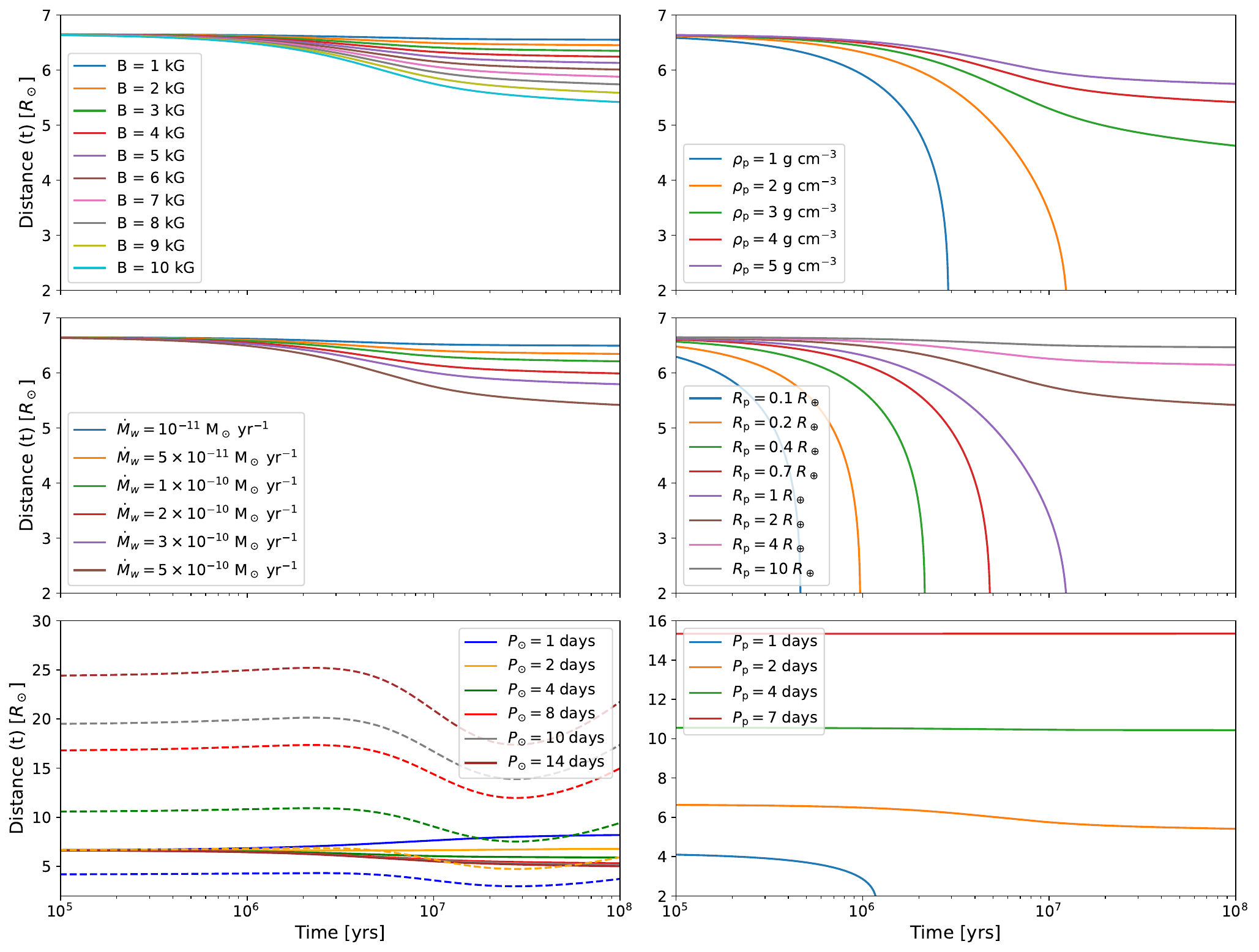}
\caption{
Orbital distance evolution of a planet under magnetic torque. In each panel, all the parameters are kept constant except one. The set of parameters used as a reference are the following: a planet density, radius, and initial period of 4 g cm$^{-3}$,  $2 R_\oplus$ and 2 days respectively (corresponding to an initial distance from the star of $\sim 3.325$ R$_{\star}$); a stellar magnetic field of 10 kG, a mass-loss rate of $5\times 10^{-10}$ M$_\odot$ yr$^{-1}$, and a rotation period of 7 days. The star's magnetic field, mass-loss rate and spin frequency are constant for $10^7$ yrs, then evolve according to equations (\ref{eq:btime}-\ref{eq:omegatime}). The left panels show the dependence of the orbital distance on stellar parameters, specifically (from top to bottom) the magnetic field, the wind mass-loss rate, and the stellar rotation period, while the right panels show the dependence of the orbital distance on the planet parameters, i.e. (from top to bottom) the planet density, radius, and orbital period).
The dashed curves in the bottom left panel represent the corotation radii.}
\label{fig7}
\end{figure*}

In this paper, we focus on the orbital evolution of fully-formed planets due to their magnetic interaction with their young or mature host stars. For clarity, we initially neglect planetary migration during the pre-main-sequence evolution of their host stars (see \S\ref{sec:ttauri} for further discussion). Given the broad range of possible initial conditions, we consider, in Figure \ref{fig7}, a wide parameter space for $B_{\rm i}$, $P_{\rm i}$, $P_{\rm p}$, $\rho_{\rm p}$, and $R_{\rm p}$ after the disk depletion.

\subsection{Early-start orbital evolution}
\label{sec:orbitalevolution}

We first consider the scenario that planets and their host stars acquired their {\it ab initio} properties at $t_{\rm start}=0$.

Figure \ref{fig:orbvspla} shows the evolution after 1 Myr, 
10 Myr and 100 Myr of the orbital distance of a planet with an initial magnetic field on the stellar surface of $B_\star =10$ kG, an initial rotation period for the star of $P_\star=7$ days, and an initial stellar-wind mass-loss rate of $5\times 10^{-10}$ M$_\odot$ yr$^{-1}$. For this fiducial model, we consider a planet with a radius of $R_{\rm p}=2 R_\oplus$, a density of $\rho_{\rm p}=4$ g cm$^{-3}$ and a mass of $M_{\rm p}=5.6 M_\oplus$. In this case, only planets located very close to the star (corresponding to an initial period $\lesssim 3$ days) experience the effects of the magnetic torque, causing their orbits to shrink and eventually leading to collisions with the stellar surface (Figure \ref{fig:orbvspla}). Planets initially located at intermediate distances (corresponding to initial periods of $\sim 3-5$ days) experience a slight reduction in their orbits, while planets located beyond the corotation radius (corresponding to initial periods of 7 days) exhibit a slight outward movement. Most of the observed evolution occurs over timescales $\lesssim 10^7$ yrs, during which, in our simplified model, the magnetic field strength,  stellar rotation period, and wind mass-loss rate decay by modest amounts over time (see equations \ref{eq:btime}-\ref{eq:omegatime}).

Based on the parameters used in Figure \ref{fig:orbvspla} as the fiducial model, we examine how magnetic torque influences the time evolution of a planet's orbit with a series of controlled experiments by varying one parameter at a time. The left panels in Figure \ref{fig7} represent quantities related to the star (from top to bottom: the stellar magnetic field, the wind mass-loss from the star, and the stellar rotation period) for a planet with $\rho_{\rm p}= 4$~g cm$^{-3}$, $R_{\rm p} = 2~R_\oplus$, and $P_{\rm p} = 2$ days, while the right panels correspond to quantities characterizing the planet (from top to bottom: the planet density, radius and initial orbital period) around a star with $B_\star (\leq 10^7 {\rm yr})=10$ kG, $M_\star=1M_\odot$, initial $R_\star=2R_\odot$, $P_\star (\leq 10^7$ yr) $=7$ days, and ${\dot M}_{\rm w} (\leq 10^7$ yr) = $5\times 10^{-10} M_\odot$ yr$^{-1}$. Across all panels, most of the evolution occurs within the time range $10^6 \;{\rm yr}<t<10^8 \;{\rm yr}$ shown in the figure.

As shown in Figure \ref{fig7}, in about half of the cases, the planet's orbit, initially located at $\sim 3.325 R_\star = 6.65 R_\odot$, shrinks, and the planet eventually collides with the stellar surface. The timescale for this event becomes shorter for increasing values of the stellar magnetic field (top, left panel), wind mass-loss rate (center left), planet period (bottom left), and for decreasing values of the planet density (top, right), planet radius (middle, right) and planet periods (bottom, right). 

With respect to the stellar magnetic field, a tenfold increase results in the shrinking of the orbit timescale by about $\sim 25\%$. The dependence on the mass-loss rate is similar.
A change in planet density by a factor of 5 (top right) leads to a more complex behavior. The orbit of planets with large densities ($\rho_p=4-5$ g cm$^{-3}$) shrink by $\sim 20-30\%$, while planets with smaller densities shrink on faster timescales and reach the stellar surface on timescales $t\gtrsim 2\times 10^{6}$ yrs. With a given $R_{\rm p}$, less dense planets have lower mass, smaller inertia, and are more affected by the magnetic drag. Likewise, a  similar effect is expected when considering the size of the planet (right-middle panel), for which a tenfold increase in the size of the planet (from 0.1 $R_\oplus$ to 1 $R_\oplus$) corresponds to an increase in the timescale for the planet to arrive to the stellar surface from $4\times 10^{5}$ yrs to  $\sim 10^{7}$ yrs.

The orbital distance is strongly influenced by both the stellar and planet periods (bottom panels). The left bottom panel shows that different outcomes are expected when the planet period (initially fixed equal to 2 days in the figure) is larger or smaller than the stellar magnetic field rotation period (the dashed lines in the bottom left panel of figure \ref{fig7}). For stellar periods $\gtrsim 4$ days, a shrinkage of the planet's orbit is expected over timescales $\gtrsim 1-2$ Myr, while an expansion of the orbit is expected if the planet is located outside the corotation radius (represented in Figure \ref{fig7} by dashed lines), as seen in the case of a planet period = 1 day. In the depicted scenario, the orbit expands from $\sim 6.6$ R$_\odot$ to $\sim 8$ R$_\star$ in about 100 Myr. 

\begin{figure}
\centering
\includegraphics[width=0.45\textwidth]{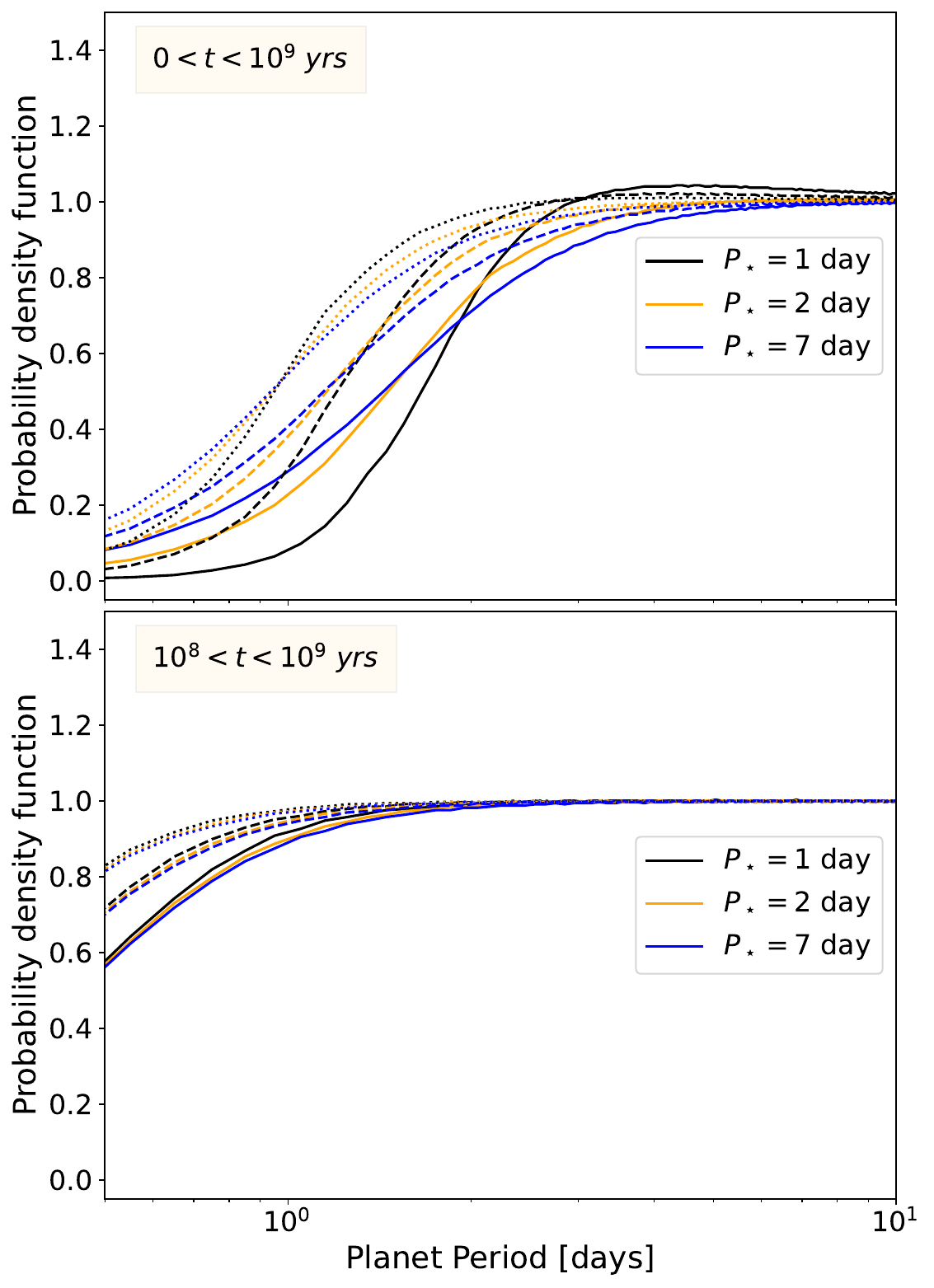}
\caption{ 
Probability density function (PDF) $\Gamma(P_{\rm p})$, giving the likelihood of finding a planet at a specific orbital period.  It is normalized to unity for all initial (at time $t_{\rm i}$) values of $P_{\rm p}$, i.e. $\Gamma (P_{\rm p}, t_{\rm i})=1$. The planet interacts with a dipolar magnetic field associated with $B_{\rm i}=10$ kG, $M_\star=1 M_\odot$, ${\dot M}_{\rm i} = 5 \times 10^{-10}M_\odot$ yr$^{-1}$, $v_{\rm w}=500$ km s$^{-1}$, and three values of $P_{\rm i}$: 1 day (black line), 2 days (orange line) and 7 days (blue line), which evolve in accordance with equations (\ref{eq:btime}-\ref{eq:omegatime}).  With $\rho_{\rm p}=4$ g cm$^{-3}$, we consider three different planetary radii, $R_{\rm p}= R_\oplus$ (solid line), $2 R_\oplus$ (dashed line), and $4 R_\oplus$ (dotted line). The upper panel represents an early-start model in which unit probability density is installed for all $P_{\rm p}$ at $t=0$.  Its distribution at $10^7$ yr shows the Alfvén-wing induced orbital evolution. 
The lower panel shows the evolution of the PDF represents a late-start model with a unit probability density installed for all $P_{\rm p}$ at $10^8$ yr. Its distribution after $10^9$ yr shows that USPs are more likely to be retained.
}
\label{fig:evo}
\end{figure}

Finally, the bottom right panel illustrates the dependence of orbit evolution on the planet period, with larger planet periods evolving over larger timescales. Consistent with the results in Figure \ref{fig:orbvspla}, planets with initial periods $\gtrsim 3$ days do not substantially alter their orbits for the set of parameters used as a reference in this figure. For a planet located close to the initial corotation radius ($P_\oplus=2$ days), the orbit remains unchanged during the first 2 Myrs. Subsequently, the stellar rotation period decreases.

The top panel of Figure \ref{fig:evo} presents the probability density function (PDF) of planet periods, $P_{\rm p}$, after $10^7$ yr of interaction with the stellar magnetic field. The PDF, $\Gamma(P_p)$, is initially flat (at time $t_{\rm i}$), with $\Gamma(P_p, t_i) = 1$. We simulate the evolution of a large number ($\sim 30000$) of planets, randomly distributed according to the periods shown in the figure. Each planet’s period is evolved over time, and the resulting period distribution is sorted into 100 bins, spanning from 0.5 s to 10 s.

We use solid, dashed, and dotted lines to represent planets with $R_p=1 R_\oplus$, $2 R_\oplus$, and $4 R_\oplus$ respectively and black, orange, and blue colors for the $P_{\rm i}=1, 2,$ and 7 days respectively.  Probabilities below unity in this plot indicate the depletion of planets with those final $P_{\rm p}$. Probabilities larger than unity correspond to the $P_{\rm p}$ range where planets accumulate beyond their {\it ab initio} population.  

While planets initially located inside the corotation radius move towards the stellar surface, planets located outside it move outwards, increasing the population of planets at intermediate periods. This effect is more pronounced for smaller stellar periods and smaller planet sizes. The severe depletion of planetary population with $P_{\rm p} \lesssim 2-3$ days is due to either their orbital decay into their host stars or their outward expansion to larger $a$ and longer $P_{\rm p}$. The greater than unity probability function in the $P_{\rm p} \sim 3.5-5$ days range is due to a pile-up of planets that migrated outwards during the early stages when the $P_\star$ was relatively small due to pre-main-sequence contraction. 

\subsection{Late-start orbital evolution}

Another possible channel for short-period planets to acquire their present-day properties is through high-$e$ migration induced by the secular perturbation of stellar or planetary companions \citep{wu2003, mardling2004}, scattering \citep{nagasawa2008}, or secular chaos \citep{wu2011}.  These processes are likely to take place long after disk depletion.  After a few $10^7$ yrs, $B_\star$ declines substantially from $B_{\rm i}$ (equation \ref{eq:btime}) and the influence of the Alfvén-wing torque remains significant only for the USPs (Figure \ref{fig:orbvspla}).

Ultrashort period planets may either be assembled {\it in situ} \citep{lee2017} or formed elsewhere and migrated to the proximity of their present-day orbits \citep{nagasawa2005, ida2010, pu2019, petrovich2019}. The former process occurs in the presence of the disk, whereas the latter mechanism can occur during disk depletion or later. Extrapolation from the results in the upper panel of Figure \ref{fig:evo} suggests that planets formed near the USPs' present-day location around T Tauri stars are unlikely to have survived.
 
We now consider the possibility that the planets relocated to the star's proximity and acquired their orbital distance $a$ after $10^8$ yr when the stellar field has decayed and $R_\star \simeq R_{\rm ms}$. In the lower panel of Figure \ref{fig:evo}, we show the probability density function of finding a planet at a specific orbital period after $10^9$ yr. All the symbols have the same definition as in the top panel. Comparison between these two panels indicates that the USP's survival probability is much higher with the delayed orbital evolution scenario. 

Based on this result, we suggest that USPs have attained their present-day orbits either around host stars with intrinsically weak stellar field throughout their lifespan or after $\sim 10^8$ yrs (Li et al., submitted) when the stellar fields have declined considerably. This also agrees with \citet{Lee25}, which show that star-planet magnetic interactions lead to ``death spiral'' and the deficit of small size ($\lesssim2R_\oplus$) planets with orbital periods shorter than $\sim 1$ day.

\section{Summary and Discussion} 
\label{sec:summary}

The omnipresence of magnetic fields on solar-type stars strongly influences on the structure and evolution of close-in hot Jupiters and close-in super-Earths, especially the ultrashort-period planets. In this paper, we consider the star-planet magnetic interaction between highly conductive planets and their host stars. Here, we carry out MHD simulations of the Alfvén-wing model \citep{strugarek2021} neglecting magnetic diffusion through the stars.  Numerical simulations of the more complex unipolar-induction process \citep{laine2012} will be presented in a subsequent paper.

While the interaction between a magnetized wind and a magnetized planet has been extensively studied, also by numerical simulations, the so-called ``unipolar'' interaction (in which a magnetized wind interacts with an unmagnetized planet) has been studied analytically \citep{laine2008,laine2012}, but not extensively by numerical simulations. The torque produced by this interaction, in particular, has been studied by \citep{strugarek2014}, who simulated both magnetized and unmagnetized planets in 2.5D (i.e. modeling the planet as a torus around the star). In such 2.5D setups, the perturbation of the magnetic field occurs primarily along the star-planet axis, whereas in our 3D simulations the perturbations extend also along the orbital direction. By performing 2D numerical simulations, we have estimated the torque exerted on an unmagnetized planet due to its interaction with the stellar magnetic field.

Then, by using a parametrization for the stellar wind mass-loss rate, stellar magnetic field, and angular momentum time evolution, we compute planetary migration across a range of stellar and planetary magnetic field configurations. While previous studies (e.g., \citealt{ahuir2021, lazovik21, lazovik23, garcia23}) have also examined planetary migration, they primarily relied on simplified torque parametrizations that were not validated by three-dimensional simulations of unmagnetized planets. Thus, our work provides a complement to previous works.

\subsection{Alfvén wings near the close-in planets.}

We show with numerical simulations (\S\ref{sec:bmach}) and analytic models (\S\ref{sec:alfvenwings}) that, in the limit of small planetary magnetic diffusivity, the relative motion between the star's unperturbed field and the planet leads to a current with an induced field in its wake. In the absence of magnetic diffusion through the planet, the pitch angle of the Alfvén wing is a function of the magnetic Mach number (Figs. \ref{fig:Bfield} and \ref{fig:diff}). The induced configuration sets a maximum conductance or minimum magnetic diffusivity (\S\ref{sec:conductance}) allowing currents to cross field lines and form a close circuit (\S\ref{sec:current}). The current density is maximum near planet's leading hemisphere (Figure \ref{fig:jfield}). Intense Ohmic dissipation in this region can cause asymmetries between ingress and egress of secondary eclipse in the light curves of stars with transiting planets (Li et al., submitted).  

Our MHD simulations are highly idealized. We plan to explore several additional boundary conditions. The effect of unipolar induction can be simulated by incorporating finite magnetic diffusivity at the boundaries of the computational domain.  We can also introduce periodic variations in $B_\star$ and $v_{\rm p}$ to simulate stars with magnetic poles misaligned with their spin axis, stars with quadrupolar or more complex unperturbed stellar magnetic fields, as well as eccentric or inclined planetary orbits.

We have neglected the intrinsic planetary magnetic field which also contributes to star-planet magnetic interactions \citep{wei2024}. \citet{strugarek2015, strugarek2016} showed that the topology of the magnetic field can change the torque by up to one order of magnitude. In the case of a planet magnetic field anti-aligned with the stellar magnetic field, in particular, the magnetic torque is about a factor of $\sim 10$ smaller than in the aligned case, where migration occurs on timescales $\gtrsim$100 Myrs. In addition, we have neglected diffusivity for the planet which likely affects the pitch angle of the Alfvén wing.  Moreover, the intense heating on the star-facing hemisphere of tidally locked USPs can lead to the formation of a magma ocean.  The magnetic diffusivity of molten silicon on the day side can be several orders of magnitude smaller than that of condensed Silicon on the night side. These effects can and will be simulated with modest modifications of our existing code.  

We have assumed a simple dipolar stellar magnetic field. In reality, the stellar magnetic field can be more complex than a simple dipolar component (e.g., \citealt{moutou16}), including quadripolar components, which have been considered in previous studies (e.g., \citealt{strugarek2015}). 
Additionally, close-in planets are irradiated by a high UV-flux from their host start, leading to atmospheric heating and potential planetary winds. These planetary wind can influence significantly the dynamical interaction between the planet and the stellar magnetic field (e.g., \citealt{matsakos15})

We have neglected thermal effects, which play a crucial role in stellar wind acceleration near the stellar surface, and can also be important in the planetary atmosphere, potentially driving planetary ``winds''. However, our simulations focus on a region close to the planet and we consider planets located inside the Alfvén radius. In this region, the wind’s magnetic pressure its typically much larger than its thermal pressure, as the (supersonic) wind speed remains lower than the local Alfvén speed (e.g., \citealt{strugarek2015}). This justifies our assumption that total pressure in the planet’s vicinity is dominated by magnetic pressure. Furthermore, \citet{strugarek2015} demonstrated that magnetic tension torques dominate over thermal pressure torques in these systems. Nevertheless, as discussed by \citet{strugarek2016}, the relative importance of thermal pressure can depend on the density profile in the lower corona.

\subsection{Initial conditions and model parameters}
\label{sec:ttauri}

We apply the torque formula to simulate the orbital evolution of close-in gas giants and super Earths. In the model-parameter study (\S\ref{sec:orbitalevolution}), we adopt a set of simple prescriptions for the evolution of host-stars' properties (equations \ref{eq:btime}-\ref{eq:omegatime}). These calculation show that: a) the co-rotation radius first moves inward during the pre-main-sequence contraction and then outward as a consequence of angular momentum loss through magnetic breaking. b) Planets with semi-major axes inside/outside the corotation radius migrate inwards/outwards. c) Planets with relatively low density, small size, and short orbital period evolve faster and are more vulnerable to migrate into their host stars.

Although our simple evolution prescriptions are qualitatively adequate during the host stars' main sequence evolution, uncertainties remain in the estimates for $B_\star$, $\rho_{\rm amb}$, and $\Omega_\star$. The initial values of these quantities are important in evaluating the direction and pace of planet's migration, when the influence of the magnetic torque is most intense.  

At advanced stages of their evolution, protoplanetary disks' accretion rate ${\dot M}_{\rm disk}$ declines on a timescale of 3-5 Myr \citep{hartmann1998} which is shorter than that for field decay, pre-main-sequence contraction, and wind-driven spin down. Protoplanets evolve into infant planets as their disk-migration winds down. Additionally, $R_{\rm cavity}$ enlarges with diminishing ${\dot M}_{\rm disk}$ (equation \ref{eq:cavitypms}). Its expansion is stalled at the inner boundary ($\sim 0.1$AU) where grains condense, the ionization fraction of the disk gas plummets, and the magnetic diffusivity increases \citep{armitage2011}. The decoupling of the stellar field with the disk can no longer maintain the $\Omega_\star-\omega_{\rm k} (R_{\rm cavity})$ synchronization. 

As host stars undergo pre-main-sequence contraction, their $R_{\rm c}$ decreases with their spin period. It is possible for protoplanets inside the magnetospheric cavity to first migrate inwards (with $a \gtrsim R_{\rm c}$) and then outwards (with $a \lesssim R_{\rm c}$) as they mature. Even at advanced stages of disk depletion, $\rho_{\rm amb}$ from the residual disk gas may largely exceed that in the stellar wind (equation \ref{eq:rhowind}). These diverse effects lead to a dispersion in the initial values of $P_\star$ and $P_{\rm p}$ in addition to the variations in $B_\star$, field decline and angular momentum loss timescales. To address these issues, a series of population-synthesis simulations will be presented elsewhere to determine the most probable  mortality rate of infant hot Jupiters and super Earths as well as the period distribution of UPSs.  

\subsection{Limited tidal torque}
\label{sec:tidaltorque}

The models presented in \S\ref{sec:migration} provide a lower-limit on the extent of migration. In addition to magnetic interaction, tidal torque also leads to the evolution of close-in planets' orbits \citep{mardling2002, dobbsdixon2004}. Contribution from the equilibrium tide \citep{hut1981, eggleton1998} is determined by the turbulent dissipation of the planet's tidal perturbation to the stellar interior \citep{zahn1977}. In the limit that the forcing frequencies induced by close-in hot Jupiters and super-Earths especially USPs, are much larger the characteristic convection-turnover frequencies, the efficiency of equilibrium tide is greatly suppressed \citep{goldreich1977, goodman1997, terquem1998}.  Dynamical tide, associated with inertial modes, can introduce frequency-dependent tidal torque \citep{ogilvie2007}.  During the pre-main-sequence contraction and during close-in planets' migration, the forcing and response frequency may pass through resonances to provide a modest averaged effective tidal torque \citep{barker2020}. Moreover, magnetic torque can induce planets to bypass the valleys of dynamical-tide \citep{wei2024}. 

However, as they age, main sequence stars lose angular momentum. When $\Omega_\star \ll \omega_{\rm p}$, the perturbing frequencies of very short-period planets, especially the UPSs, fall outside the inertial range for dynamical tide to be excited \citep{ogilvie2014}. Similar to the magnetic torque, the magnitude of the tidal torque decreases with time and planets' $a$ and the close-in planets' orbital evolution occurs mostly around young stars. This expectation is consistent with the correlation between the occurrence rate of hot Jupiters and stellar age, as inferred from GAIA data \citep{hamer2019}. The present-day equilibrium and dynamical tidal torque exerted on the USPs (with $\Omega_\star \ll \omega_{\rm P}$) are expected to be much weaker than the magnetic torque induced by weakly active main sequence stars.  We also infer from the lower panel of Figure \ref{fig:evo} that they relocated to the proximity of their present orbits after the magnetic field of their host stars have decayed substantially on the timescale of $\sim 10^8$ yr. 

Comparison between tidal and magnetic torques \citep{strugarek2017, wei2024} throughout the host star's pre-main-sequence and main-sequence evolution will be presented in a subsequent paper. A population synthesis study can provide useful predictions on the evolution of period distribution of close-in planets and the occurrence rate of USPs \citep{ahuir2021}.


\begin{figure}
\centering
\includegraphics[width=0.45\textwidth]{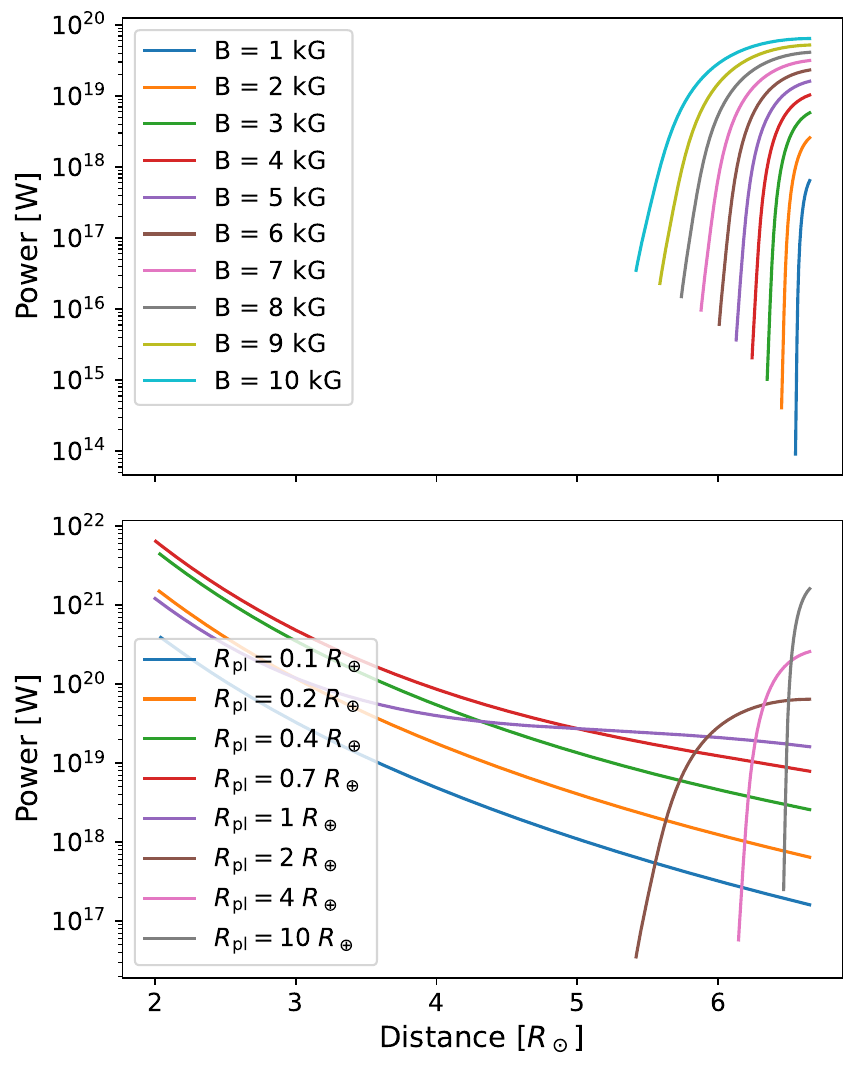}
\caption{Radio luminosity emitted during the migration of  a planet under magnetic torques. The set of parameters used as a reference are the same used in figure \ref{fig7}: a planet density, radius and initial period of 4 g cm$^{-3}$,  $2 R_\oplus$ and 2 days respectively (corresponding to an initial distance from the star of $\sim 3.325$ R$_{\star}$); a stellar magnetic field of 10 kG, a mass-loss rate of $5\times 10^{-10}$ M$_\odot$ yr$^{-1}$, and a rotation period of 7 days. The stellar magnetic field, the mass-loss rate and the rotation period are constant for $10^7$ yrs, then evolve according to equations (\ref{eq:btime}-\ref{eq:jtime}). The upper and bottom panels show the dependence of the power on the stellar magnetic field and on the planet radius.}
\label{fig:radio}
\end{figure}

\subsection{Radio emission from the Alfvén wing}

One potentially observable signature of the Alfvén wing model is its associated radio emission. Charged particles along the distorted field lines emit cyclotron radiation in the radio frequency range. This process occurs in both the stellar field perturbed by USPs and in the planets' field perturbed by their moons \citep{goldreich1969}. The power of radio emissions due to the Alfvén wing interaction can be estimated using the radio-magnetic Bode's law \citep{zarka2007}
\begin{equation}\label{power}
  W_{\rm \star p, p m} \approx \eta_{\rm \star p, p m} v_{\rm \star p, p m} B_{\rm \star p, pm}^2 R_{\rm \star p, pm}^2 \pi/\mu
\end{equation}
Finally, we note that, while Bode's law has been used extensively, \citet{nichols11} pointed out that the current system is inherently limited by the Alfvén conductance, and therefore can hardly support the large currents required by the radio-magnetic Bode’s law in the parameter range of hot exoplanets. Following \citet{nichols11}, computing properly the radio emission could require considering the complex magnetosphere–ionosphere coupling.

\section*{Acknowledgements}

We thank Randy Laine, Andrew Cumming, Xing Wei, Lee Hartmann, Moira Jardine, Kevin Schlaufman, Eve Lee and Enrico Ramirez-Ruiz for useful conversations. We thank the anonymous referee for suggestions which helped improving substantially the paper. FDC acknowledges support from the DGAPA/PAPIIT grant IN113424. We acknowledge the computing time granted by DGTIC UNAM on the supercomputer Miztli (project LANCAD-UNAM-DGTIC-281). 

\section*{Data Availability}

The data underlying this article will be shared on reasonable request to the corresponding author.

\bibliography{bibliography}{}
\bibliographystyle{mnras}

\section*{Appendix}
\label{appendix}

\begin{figure}
\centering
\includegraphics[width=0.45\textwidth]{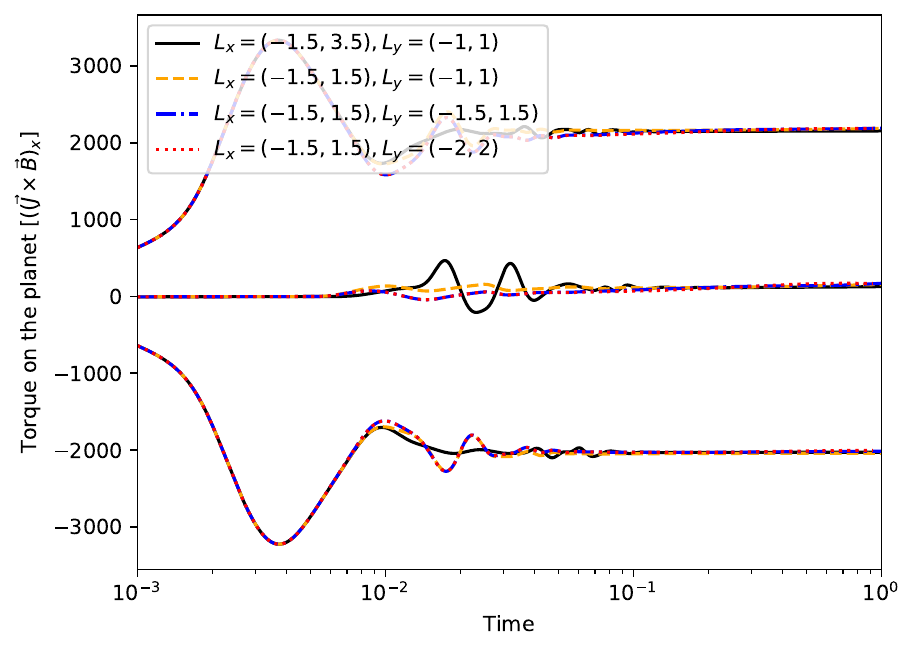}
\caption{
Force produced by the interaction of the magnetic field with the planet, for different sizes of the computational region. Except for small fluctuations present at intermediate times in the simulation with larger computational box along the $x$ axis (solid, black line), the results converge to the same values in all cases. The simulations correspond to an initial magnetic field $B_0=100$. The groups of curves correspond to: $-0.5<x-x_p< 0$ (upper curves), $0.5<x-x_p< 0.5$ (bottom curves), and the rest of the computational volume (curves with $F\sim 0$). 
}
\label{fig10}
\end{figure}

\begin{figure}
\centering
\includegraphics[width=0.45\textwidth]{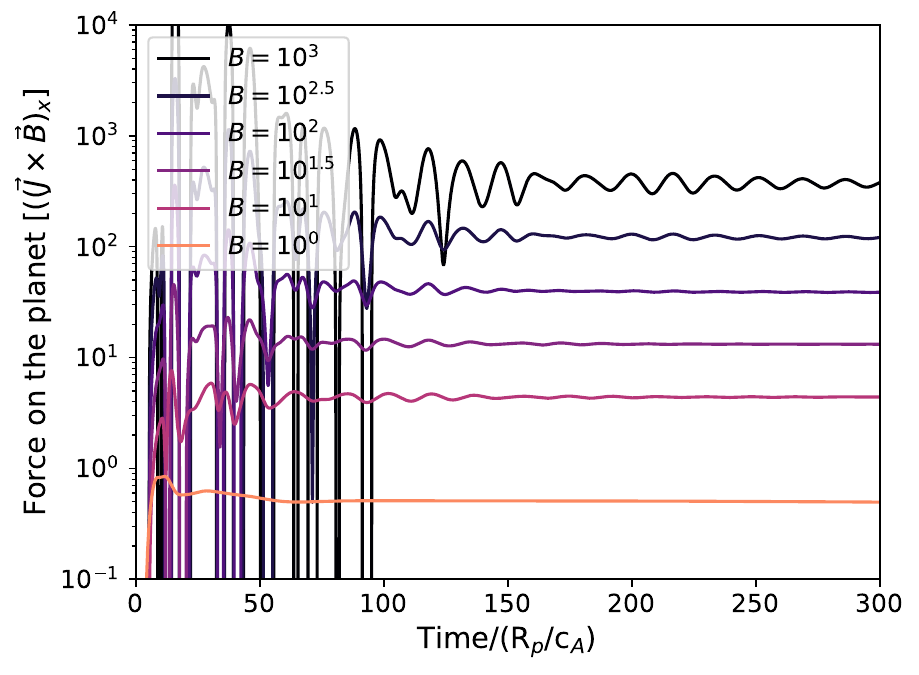}\\
\includegraphics[width=0.45\textwidth]{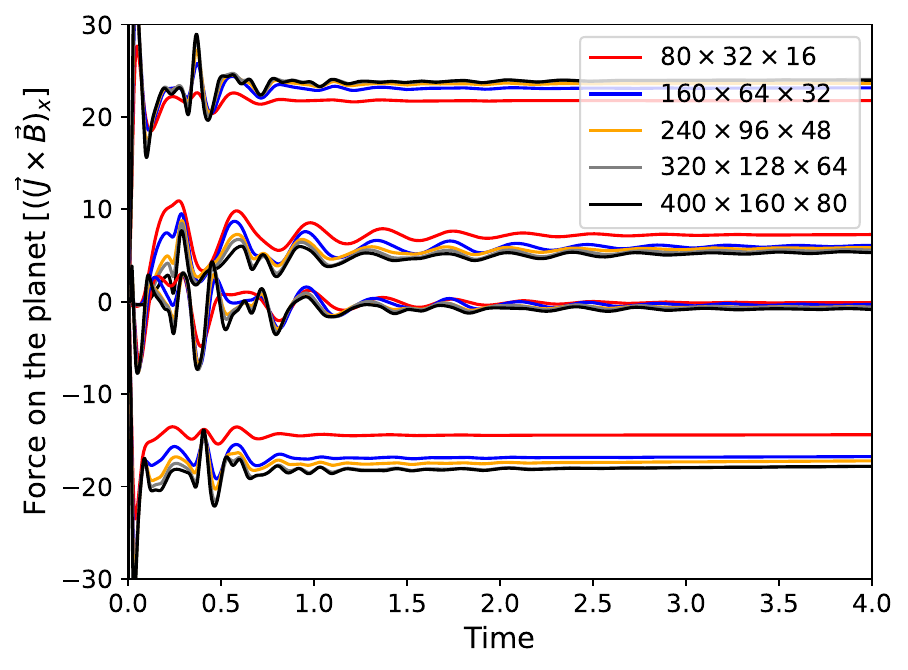}
\caption{
\emph{Top panel}: Time evolution of the magnetic force for different stellar magnetic fields, ranging from $B_0=1$ and $B_0=10^3$. In these simulations, $\eta_{\rm p} = 0$, and $\eta_{\rm amb}= 10^{-5}$. The time evolution indicates convergence toward asymptotic values. 
\emph{Bottom panel}: Magnetic force produced by the interaction of the magnetic field with the planet, for a stellar magnetic field of $B_0=10$, shown for different resolutions and computational domains. 
The groups of curves correspond to: $-0.5<x-x_p< 0$ (upper curves), $0.5<x-x_p< 0.5$ (bottom curves), the rest of the computational volume (curves with $F\sim 0$), and the entire computational domain (curves with $5\lesssim F \lesssim 10$) . 
}
\label{fig:res}
\end{figure}

To minimize the presence of numerical artifacts in the simulation results, we examined how the results depend on the size of the computational box and the resolution employed. Figures \ref{fig10} and \ref{fig:res} show that the calculation of the force exerted by the magnetic field on the planet (used to compute the evolution of the orbital separation) does not strongly depend on either the size of the box or the numerical resolution employed. 
The top panel of Figure \ref{fig:res} also shows that simulations with different magnetic field strengths converge to the asymptotic steady-state solution on very different timescales, with simulations with larger magnetic fields converging much faster. A magnetic field line travels a distance $R_{\rm p}$ in a time $\sim R_{\rm p}/v_{\rm k} \sim \sqrt{a R_{\rm p}^2/GM_\star} \lesssim 100$ s, where $M_\star$ is the mass of the star, and $a$ is the orbital semi major axis of the planet (\S\ref{sec:codeunit}). This timescale is negligible compared to the evolution timescale of the system ($\sim$ several million years). Around the planet, the magnetic field lines move along the $z$ direction on a timescale $\sim R_{\rm p}/c_{\rm B} \propto 1/\sqrt{B_0}$ (see above discussions) and it generally differs from the planet's travel timescale.

\bsp	
\label{lastpage}

\end{document}